\documentclass[aps,prl,reprint,superscriptaddress,floatfix]{revtex4-2}

\usepackage{graphicx}
\usepackage{amsmath,amssymb,amsfonts,bm}
\usepackage{amsthm}
\usepackage{mathrsfs}
\usepackage[dvipsnames,x11names]{xcolor}
\usepackage{booktabs}
\usepackage{siunitx}
\usepackage{dcolumn}
\usepackage{multirow}
\usepackage{braket}
\usepackage[normalem]{ulem}
\usepackage{textcomp}
\usepackage{listings}
\usepackage{soul}
\usepackage{comment}
\usepackage{algpseudocode}
\usepackage{ragged2e}

\newcommand{\Msun}{M_{\odot}}
\renewcommand{\.}{\ensuremath{\mspace{1mu}}}
\renewcommand{\arraystretch}{1.5}

\begin{document}

\title{Gravitational Imprints of Axion Clouds on Black-Hole Orbital Dynamics}

\author{J.~Ruz}
\thanks{Corr. author: \mbox{Jaime.Ruz@tu-dortmund.de}}
\affiliation{ Fakult{\"a}t f{\"u}r Physik,
Technische Universit{\"a}t Dortmund,
44227 Dortmund, Germany}
\affiliation{
Cluster of Excellence “Color meets Flavor”, Technische Universit{\"a}t Dortmund}

\author{V.~Mungai}
\thanks{Corr. author: \mbox{Veronica.Mungai@tu-dortmund.de}}
\affiliation{
Fakult{\"a}t f{\"u}r Physik,
Technische Universit{\"a}t Dortmund,
44227 Dortmund, Germany}
\affiliation{
Cluster of Excellence “Color meets Flavor”, Technische Universit{\"a}t Dortmund}

\author{F.~K{\"u}hnel}
\thanks{Corr. author: \mbox{Florian.Kuehnel@physik.uni-muenchen.de}}
\affiliation{
    Fakult{\"a}t f{\"u}r Physik,
    Technische Universit{\"a}t Dortmund,
    44227 Dortmund,
    Germany}
\affiliation{
	Arnold Sommerfeld Center,
	Ludwig-Maximilians-Universit{\"a}t,
	Theresienstr.~37,
	80333 M{\"u}nchen,
	Germany}

\author{C.~K{{\"o}}hn}
\email{Christoph.Koehn@tu-dortmund.de}
\affiliation{
Fakult{\"a}t f{\"u}r Physik,
Technische Universit{\"a}t Dortmund,
44227 Dortmund, Germany}
\affiliation{ Department of Space Physics and Technology,
Technical University of Denmark,
2800 Kongens Lyngby, Denmark}

\author{E.~Stamou}
\email{Emmanuel.Stamou@tu-dortmund.de}
\affiliation{
Fakult{\"a}t f{\"u}r Physik,
Technische Universit{\"a}t Dortmund,
44227 Dortmund, Germany}
\affiliation{
Cluster of Excellence “Color meets Flavor”, Technische Universit{\"a}t Dortmund}

\author{J.~K.~Vogel}
\email{Julia.Vogel@tu-dortmund.de}
\affiliation{
Fakult{\"a}t f{\"u}r Physik,
Technische Universit{\"a}t Dortmund,
44227 Dortmund, Germany}
\affiliation{
Cluster of Excellence “Color meets Flavor”, Technische Universit{\"a}t Dortmund}

\author{B.~Guichard}
\email{baptiste.guichard@tu-dortmund.de}
\affiliation{
Fakult{\"a}t f{\"u}r Physik,
Technische Universit{\"a}t Dortmund,
44227 Dortmund, Germany}


\date{\today}

\begin{abstract}
Black holes of primordial or astrophysical origin immersed in axion dark matter provide a natural laboratory for exploring the interplay between compact objects and ultralight scalar fields. While previous studies have primarily focused on superradiance and horizon-scale phenomena, the gravitational response of the surrounding axion distribution has received comparatively little attention. We investigate the dynamics of particles orbiting Schwarzschild and Kerr primordial black holes embedded in scalar-field dark matter halos. Using first-order perturbation theory, we show that the ambient halo redistributes the gravitationally bound axion cloud, shifting the characteristic radius of the gravitational atom. Numerical integrations of timelike geodesics demonstrate that the maximum orbital response consistently occurs at this perturbed radius, establishing a direct correspondence between the microscopic structure of the axion wavefunction and the macroscopic dynamics of orbiting particles. Our results identify a new purely gravitational signature of scalar-field dark matter around primordial black holes and provide a framework for probing mixed axion--primordial black hole dark matter scenarios.
\end{abstract}

\maketitle

\noindent\textit{Introduction\,---\,}The nature of dark matter remains one of the foremost open questions in modern physics. Among the leading non-baryonic dark matter candidates, axions \cite{Peccei:1977hh, Peccei:1977ur, Weinberg:1977ma, Wilczek:1977pj} and primordial black holes (PBHs) \cite{Hawking:1971ei, Carr:1974nx, Carr:1975qj} are particularly compelling because both are non-thermal relics of the early Universe whose cosmological abundances are determined by early-Universe dynamics rather than thermal freeze-out. As such, they provide complementary probes of inflation, Peccei--Quinn symmetry breaking in the axion sector, and primordial gravitational collapse~\cite{Marsh:2015xka, DiLuzio:2020wdo, Escriva:2022duf, Carr:2020xqk, Carr:2016drx}.

Although axions and PBHs are often studied independently, they are not mutually exclusive. Mixed dark-matter scenarios containing both components arise naturally in several cosmological models, including PBH formation in axion cosmologies and axion production during PBH-dominated eras \cite{Bernal:2021bbh, Domcke:2017fix, Ferrer:2018uiu}. In such scenarios, every PBH is expected to reside within an ambient axion dark-matter distribution, providing a natural environment in which compact objects and ultralight bosonic fields coexist.

The interaction between axions and black holes has been extensively studied through black-hole superradiance and gravitational atoms~\cite{Arvanitaki:2010sy, Brito:2015oca, Baryakhtar2021}, while scalar-field dark matter (SFDM) halos have been investigated as astrophysical realizations of ultralight bosonic matter and for their gravitational influence on compact objects~\cite{Hu2000, Hui2017, Suarez2014, Ferreira2021}. However, the connection between the environmental redistribution of gravitationally bound axion states and observable orbital dynamics has not been established.

\begin{figure}[!b]
    \hspace{-0.5 cm}
    \includegraphics[width=1.06\linewidth]{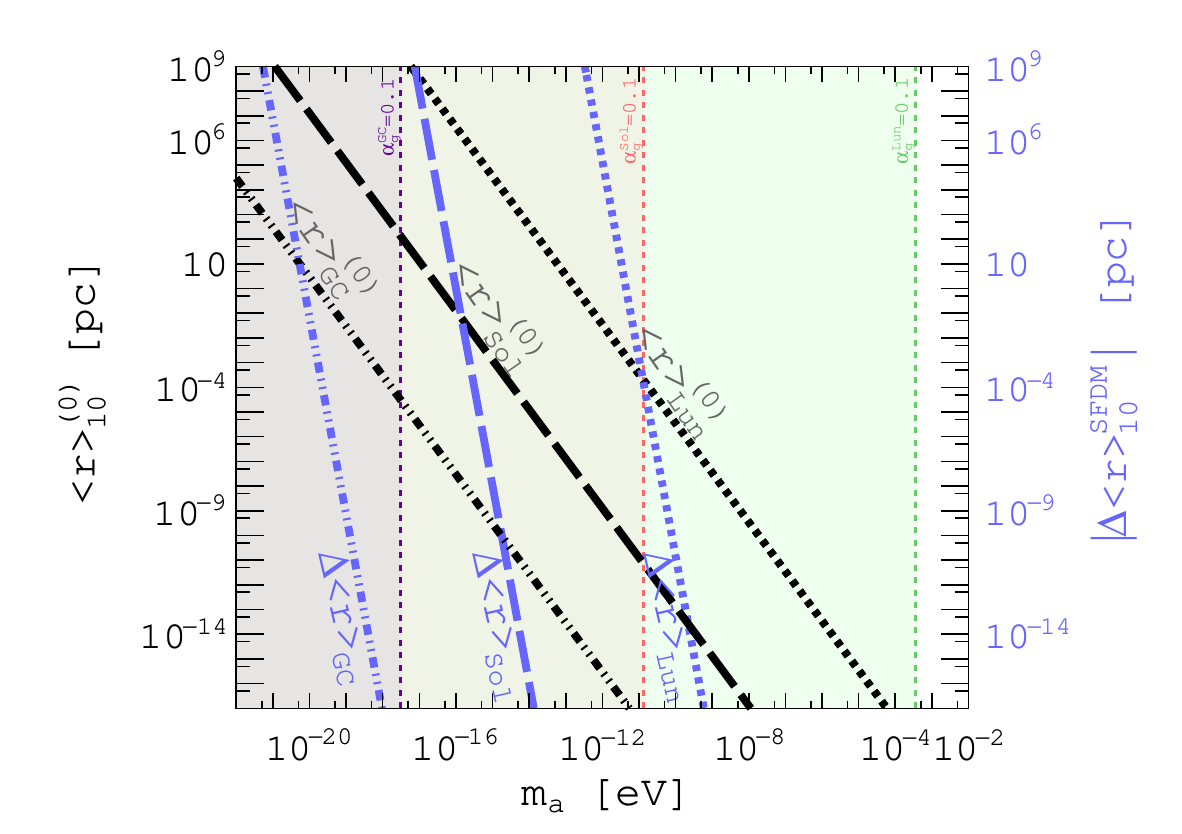}
    \caption{
Unperturbed expectation radius $\langle r\rangle_{10}^{(0)}$ (black) and
magnitude of the SFDM-induced correction
$|\Delta\langle r\rangle_{10}^{\rm SFDM}|$ (blue) as functions of the
axion mass $m_a$, for lunar-mass
($M_{\rm BH}=3.7\times10^{-8}M_\odot$), solar-mass
($M_{\rm BH}=M_\odot$), and Galactic-Center
($M_{\rm BH}=4.3\times10^{6}M_\odot$) black holes.
Different line styles distinguish the three black-hole masses.
Vertical dashed lines mark
$\alpha_g\equiv G M_{\rm BH}m_a=0.1$ for each benchmark, with the
corresponding shaded regions to the left indicating $\alpha_g<0.1$,
i.e., the weak-coupling regime
The SFDM correction is evaluated in the compact-cloud limit,
$\Delta\langle r\rangle_{10}^{\rm SFDM}
\simeq -9\pi m_a^2\rho_c a_g^5$.
}
\label{fig:delta-r-pert}
\end{figure}

In this {\it Letter}, we investigate test-particle motion around Schwarzschild and Kerr primordial black holes embedded in axion
Bose--Einstein-condensate halos. The surrounding scalar-field halo acts as a weak gravitational perturbation of the bound axion states, redistributing their hydrogenic wavefunctions while preserving the
underlying quantum-number structure. 

Figure~\ref{fig:delta-r-pert} summarizes the central perturbative prediction of this {\it Letter}. The surrounding scalar-field halo redistributes the hydrogenic wavefunctions of gravitationally bound axions, shifting the vacuum expectation value $\langle r\rangle_{n\ell}^{(0)}$  to the perturbed cloud radius \mbox{$\langle r\rangle_{n\ell}^{\rm SFDM} = \langle r\rangle_{n\ell}^{(0)} + \Delta\langle r\rangle_{n\ell}^{\rm SFDM}$}. Within the hydrogenic approximation, this radius provides a direct prediction for where the gravitational influence of the cloud on orbital motion should be maximal. Numerical integrations of timelike geodesics confirm this prediction, demonstrating that the characteristic orbital-response radius coincides with the perturbatively predicted cloud radius. This correspondence establishes a direct connection between the quantum structure of gravitationally bound axion clouds and the classical dynamics of orbiting particles, revealing a new purely gravitational signature of mixed axion--PBH dark matter.

\vspace{0.2cm}
\noindent\textit{Black Holes Embedded in Scalar-Field Dark Matter\,---\,}In order to model the gravitational environment surrounding a  black hole, we embed the vacuum spacetime in an equilibrium scalar-field dark-matter (SFDM) halo. Following matter--radiation equality, primordial black holes naturally evolve within the gravitational potential of their host dark-matter halos. We model the surrounding halo as SFDM, motivated by ultralight axions whose large occupation numbers admit a coherent Bose--Einstein condensate (BEC) description governed by the coupled Schr\"odinger--Poisson equations~\cite{Hu2000,Sikivie:2009qn,Suarez2014,Hui2017,Ferreira2021}. To isolate the gravitational response of gravitationally bound axion states, we adopt the Thomas--Fermi equilibrium profile,
\begin{equation}
    \rho_{\rm BEC}(r)
        =
            \rho_{\rm c}
            \frac{\sin(\pi\.r/R)}
            {\pi\.r/R}
            \, ,
            \label{eq:BECprofile}
\end{equation}
where $\rho_{\rm c}$ denotes the central density and $R$ the condensate radius, defined by the first zero of the density profile. Throughout this work, the halo is treated as a fixed equilibrium background, neglecting its dynamical response to the central black hole.

The Schwarzschild spacetime embedded in the scalar-field halo is described by~\cite{Xu:2018}
\begin{equation}
    {\rm d}s^{2}
        =
            -f(r)\,{\rm d}t^{2}
            +
            f(r)^{-1}{\rm d}r^{2}
            +
            r^{2}
            \big(
                {\rm d}\theta^{2}+\sin^{2}\theta\,{\rm d}\phi^{2}
            \big)
            \, ,
            \label{eq:metric-general}
\end{equation}
with
\begin{equation}
    f(r)
        =
            \exp\big[ -\Phi_h(r) \big]
            -
            \frac{ 2\.M_{\rm BH} }{ r }
            \, ,
            \label{eq:fmetric}
\end{equation}
where
\begin{equation}
    \Phi_h(r)
        =
        \frac{8R^{2}}{\pi}\rho_{\rm BEC}(r)
            \, .
\end{equation}
Here $M_{\rm BH}$ denotes the black-hole mass and geometrised units ($G=c=1$) are assumed throughout.

The rotating solution follows from~\cite{Xu:2018}. Using an equivalent notation convenient for the present analysis, we define
\begin{equation}
    \Sigma
        \equiv
            r^{2} + a^{2}\.\cos^{2}\theta
            \, 
\end{equation}
and
\begin{equation}
    \Delta
        \equiv
            r^{2}\.
            \exp\!\big[ -\Phi_h(r) \big]
            -
            2\.M_{\rm BH}\.\.r
            +
            a^{2}
            \, ,
\end{equation}
the Kerr line element becomes
\begin{align}
    {\rm d}s^{2}
        =&
            -
            \Bigg[
                1-
                \frac{
                r^{2}+2\.M_{\rm BH}\.\.r-r^{2}\.e^{-\Phi_h(r)}
                }{ \Sigma }
            \Bigg]\.
            {\rm d}t^{2}
            +
            \frac{\Sigma}{\Delta}{\rm d}r^{2}
            \nonumber
            \\[2 mm]
    &
            +
            \frac{\sin^{2}\theta}{\Sigma}
            \left[
              \big( r^{2} + a^{2} \big)^{2}
                -
                a^{2}\Delta\sin^{2}\theta
            \right]{\rm d}\phi^{2}
            +
            \Sigma\,{\rm d}\theta^{2}
            \nonumber
            \\[2 mm]
    &
            -2
            \left[
                r^{2}+2\.M_{\rm BH}\.\.r-r^{2}e^{-\Phi_h(r)}
            \right]
            \frac{a\sin^{2}\theta}{\Sigma}
            \,{\rm d}t\,{\rm d}\phi
            \, .
\end{align}

The black-hole rotation is characterised by the Kerr spin parameter $a \equiv J/M_{\rm BH}$, where $J$ is the angular momentum. The vacuum Schwarzschild and Kerr metrics are recovered in the limit $\rho_{\rm c} \rightarrow 0$. Throughout this work, identical initial conditions are evolved in both vacuum and halo spacetimes, allowing differences in the resulting trajectories to be attributed solely to the gravitational field of the surrounding axion condensate.

\vspace{0.2cm}
\noindent\textit{Perturbative Redistribution of the Axion Cloud\,---\,}
We distinguish between the ambient SFDM halo and the
black-hole-bound axion cloud. The former is treated as an externally
prescribed, spherically symmetric gravitational background, whereas the
latter corresponds to a gravitationally bound state of the black hole.
We focus on the regime in which the black-hole potential dominates the
bound-state dynamics and the surrounding halo acts as a weak gravitational
perturbation. In the weak-field and non-relativistic limits, its contribution
to the bound-state Hamiltonian is
\begin{equation}
    \delta V(r)=m_a\Psi_{\rm SFDM}(r),
\end{equation}
where $\Psi_{\rm SFDM}$ is the Newtonian potential generated by the
ambient halo.

The axion field $\Phi$ satisfies the covariant Klein--Gordon equation
\begin{equation}
    \left( \Box - m_{\rm a}^{2} \right)\Phi
        =
            0
            \, ,
            \label{eq:KG}
\end{equation}
where $m_{\rm a}$ is the axion mass. In the weak-field regime,
\begin{equation}
    g_{tt}
        \simeq
            -(1+2\Psi)
            \, ,
\end{equation}
with
\begin{equation}
    \Psi(r)
        =
            \Psi_{\rm BH}(r)
            +
            \Psi_{\rm SFDM}(r)
            \, ,
\end{equation}
where $\Psi_{\rm BH}(r) = -M_{\rm BH}/r$ and
\begin{equation}
    \Psi_{\rm SFDM}(r)
        =
            -\frac{ 1 }{ 2 }\,\Phi_h(r)
        =
            -\frac{4\.\.\rho_{\rm c}R^2}{\pi}
            \frac{\sin(\pi\.r/R)}{\pi\.r/R}
            \, .
\label{eq:Psi_halo}
\end{equation}

Writing the relativistic field in terms of a slowly varying
non-relativistic mode,
\begin{equation}
\Phi(t,\mathbf{x})
=
\frac{1}{\sqrt{2m_{\rm a}}}
\psi(t,\mathbf{x})e^{-im_{\rm a}t},
\end{equation}
where $\psi$ varies slowly compared with the rest-mass oscillation, the
non-relativistic reduction of Eq.~(\ref{eq:KG}) yields

\begin{equation}
    i\frac{\partial\psi}{\partial t}
        =
            \left[
                -\frac{\nabla^2}{2\.m_{\rm a}}
                +
                m_{\rm a}\Psi(r)
            \right]\psi
            \, ,
\label{eq:Schrodinger}
\end{equation}
whose derivation is given in the {\it Supplemental Material}. In the weak-coupling regime,
\begin{equation}
    \alpha_{\rm g}
        \equiv
            M_{\rm BH}\.\.m_{\rm a}
        \ll
            1
            \, ,
\end{equation}
the Hamiltonian separates naturally into
\begin{equation}
    H
        =
            H_{0} + \delta H_{\rm SFDM}
            \, ,
            \label{eq:H}
\end{equation}
where
\begin{equation}
    H_{0}
    =
        -\frac{\nabla^2}{2\.m_{\rm a}}
        -
        \frac{M_{\rm BH}\.\.m_{\rm a}}{r}
        \, ,
\label{eq:H0}
\end{equation}
is the gravitational analogue of the hydrogen atom~\cite{Detweiler:1980, Dolan:2007mj, Arvanitaki:2010sy, Brito:2020lpr}, and
\begin{equation}
    \delta H_{\rm SFDM}
        =
            m_{\rm a}\.\Psi_{\rm SFDM}(r)
        =
            -\frac{4\.m_{\rm a}\.\.\rho_{\rm c}\.R^2}{\pi}\.\.
            \frac{\sin(\pi\.r/R)}{\pi\.r/R}
            \, .
            \label{eq:dH}
\end{equation}
The unperturbed bound states are characterised by the gravitational Bohr radius
\begin{equation}
    a_{\rm g}
        =
            \frac{1}{M_{\rm BH}\.\.m_{\rm a}^2}
            \, ,
\end{equation}
and the hydrogenic expectation value
\begin{equation}
    \langle r\rangle_{n\ell}^{(0)}
        =
            \frac{a_{\rm g}}{2}
            \left[
                3n^2-\ell(\ell+1)
            \right]
            ,
\label{eq:rvac}
\end{equation}
which defines the characteristic size of the gravitational atom. For the ground state considered below, $\langle r\rangle_{10}^{(0)}=3a_g/2$.

Since our goal is to relate the microscopic cloud structure to macroscopic orbital dynamics, the relevant observable is the perturbation of the wavefunction rather than the energy spectrum. To first order,
\begin{equation}
    \Big| \psi_{n\ell m}^{\rm SFDM} \Big\rangle
        =
            |n\ell m\rangle
            +
            \sum_{n'\neq n}
            \frac{ \big\langle n'\ell m \big| \delta H_{\rm SFDM} \big| n\ell m \big\rangle}{E_n^{(0)}-E_{n'}^{(0)}}\,
            \big| n'\ell m \big\rangle
            \, ,
\label{eq:dpsi}
\end{equation}
where $E_n^{(0)} = -m_{\rm a}\alpha_{\rm g}^2/(2n^2)$ are the unperturbed binding energies. Owing to spherical symmetry, the perturbation preserves $\ell$ and $m$, so that only states with different $n'$ contribute to the radial mixing.

The corresponding shift of the characteristic cloud radius is
\begin{equation}
    \langle r\rangle_{n\ell}^{\rm SFDM}
        =
            \langle r\rangle_{n\ell}^{(0)}
            +
            \Delta\langle r\rangle_{n\ell}^{\mathrm{SFDM}}
            \, ,
\label{eq:rshift}
\end{equation}
with
\begin{equation}
    \Delta\langle r\rangle_{n\ell}^{\mathrm{SFDM}}
        =
            2\,{\rm Re}
            \sum_{n'\neq n}
            \frac{\mathcal V^{(\ell)}_{n'n}\mathcal R^{(\ell)}_{nn'}}
            {E_n^{(0)}-E_{n'}^{(0)}}
            \, .
            \label{eq:deltaR}
\end{equation}
Here $\mathcal R^{(\ell)}_{nn'}=\langle n\ell m|r|n'\ell m\rangle$, while the explicit radial matrix elements $\mathcal V^{(\ell)}_{n'n}$ are given in the {\it Supplemental Material}. Equation~(\ref{eq:deltaR}) is the central analytical result of this work: the scalar-field halo perturbs the spatial extent of the gravitational atom, suggesting that the strongest orbital response should occur near the perturbed expectation value $\langle r\rangle_{n\ell}^{\rm SFDM}$.

For compact clouds satisfying $a_{\rm g}\ll R$, the leading-order corrections reduce to
\begin{equation}
    \Delta E_{10}^{\rm SFDM}
    \simeq
        -\frac{4m_{\rm a}\rho_{\rm c}R^2}{\pi}
        \, ,
    \qquad
    \Delta\langle r\rangle_{10}^{\rm SFDM}
        \simeq
            -9\pi m_{\rm a}^2\rho_{\rm c}a_{\rm g}^5
            \, ,
\end{equation}
demonstrating that the halo compresses the ground-state wavefunction.

Figure~\ref{fig:delta-r-pert} summarizes the characteristic cloud size and its perturbative correction across the parameter space. The unperturbed radius follows the hydrogenic scaling $\langle r\rangle_{10}^{(0)} \propto (M_{\rm BH}\.\.m_{\rm a}^2)^{-1}$, whereas $|\Delta\langle r\rangle_{10}^{\mathrm{SFDM}}| \propto (M_{\rm BH}^5m_{\rm a}^8)^{-1}$, indicating that the largest perturbative effects arise for light axions bound to low-mass primordial black holes.

In the following section we test this prediction using numerical integrations of timelike geodesics in the full Schwarzschild and Kerr spacetimes. As we shall show, the radius of maximum orbital response closely follows the perturbatively predicted value of $\langle r\rangle_{n\ell}^{\rm SFDM}$, establishing a direct correspondence between the quantum structure of gravitationally bound axion clouds and classical orbital dynamics.

\vspace{0.2cm}

\noindent\textit{Orbital Response and Numerical Validation\,---\,}To test the perturbative prediction, we numerically integrate timelike geodesics~\cite{Bacchini:2018, Cristoph} in both the vacuum and scalar-field dark-matter (SFDM) spacetimes, using identical initial positions and four-velocities. The code used for the black-hole--halo geodesic simulations is publicly available on GitHub~\cite{NumericalRelativity}. The particle motion satisfies
\begin{equation}
    \frac{{\rm d}^{2}x^\mu}{{\rm d}\tau^{2}}
    +
    \Gamma^\mu_{\alpha\beta}
    \frac{{\rm d}x^\alpha}{{\rm d}\tau}
    \frac{{\rm d}x^\beta}{{\rm d}\tau}
        =
            0
            \, ,
            \label{eq:geodesic}
\end{equation}
where the Christoffel symbols are computed from either the vacuum Schwarzschild/Kerr metric or the corresponding spacetime embedded in an SFDM halo. 

We consider both Schwarzschild black holes and slowly rotating Kerr spacetimes with representative dimensionless spin parameters $a_\ast\leq0.2$, thereby capturing the leading effects of black-hole rotation while remaining in the weak-spin regime. To demonstrate the scalability of the framework across vastly different mass and environmental scales, we consider five benchmark configurations. The first three correspond to compact SFDM clouds surrounding black holes of masses $M_{\rm BH}=1\,M_\odot$, $10\,M_\odot$, and $4.3\times10^{6}\,M_\odot$, respectively. For the two stellar-mass benchmarks, we choose $m_{\rm a}=4.28\times10^{-11}\,\mathrm{eV}$ and $4.28\times10^{-12}\,\mathrm{eV}$, respectively, corresponding to $\alpha_g\simeq0.32$ in both cases. For the $4.3\times10^{6}\,M_\odot$ compact-cloud benchmark, we take $m_{\rm a}=10^{-17}\,\mathrm{eV}$. The remaining two configurations describe supermassive black holes embedded in extended galactic SFDM halos. The first represents the Galactic Centre, modelled with $M_{\rm BH}=4.3\times10^{6}\,M_\odot$~\cite{GRAVITY_2022MassDistribution} and $m_{\rm a}=10^{-17}\,\mathrm{eV}$, while the second corresponds to the central black hole of ESO~120-0211, with $M_{\rm BH}=5.62\times10^{6}\,M_\odot$ and $m_{\rm a}=7.6\times 10^{-18}\,\mathrm{eV}$. Further details of the cloud and halo parameters for these benchmark configurations are provided in the Supplemental Material.

Although the halo only weakly modifies the overall orbital morphology, the radial deviation from the corresponding vacuum trajectory develops a pronounced maximum within a finite radial interval. Inside this region the halo systematically contracts the orbit, $r_{\rm SFDM} < r_{\rm vac}$, while simultaneously inducing an accumulated azimuthal shift. Outside this interval both effects rapidly disappear, indicating that the gravitational influence of the halo is localised rather than uniformly distributed along the orbit.

\begin{figure}[!t]
    \centering
    \begin{minipage}{0.49\textwidth}
    \hspace{-.75 cm}
         \includegraphics[width=0.9\textwidth]{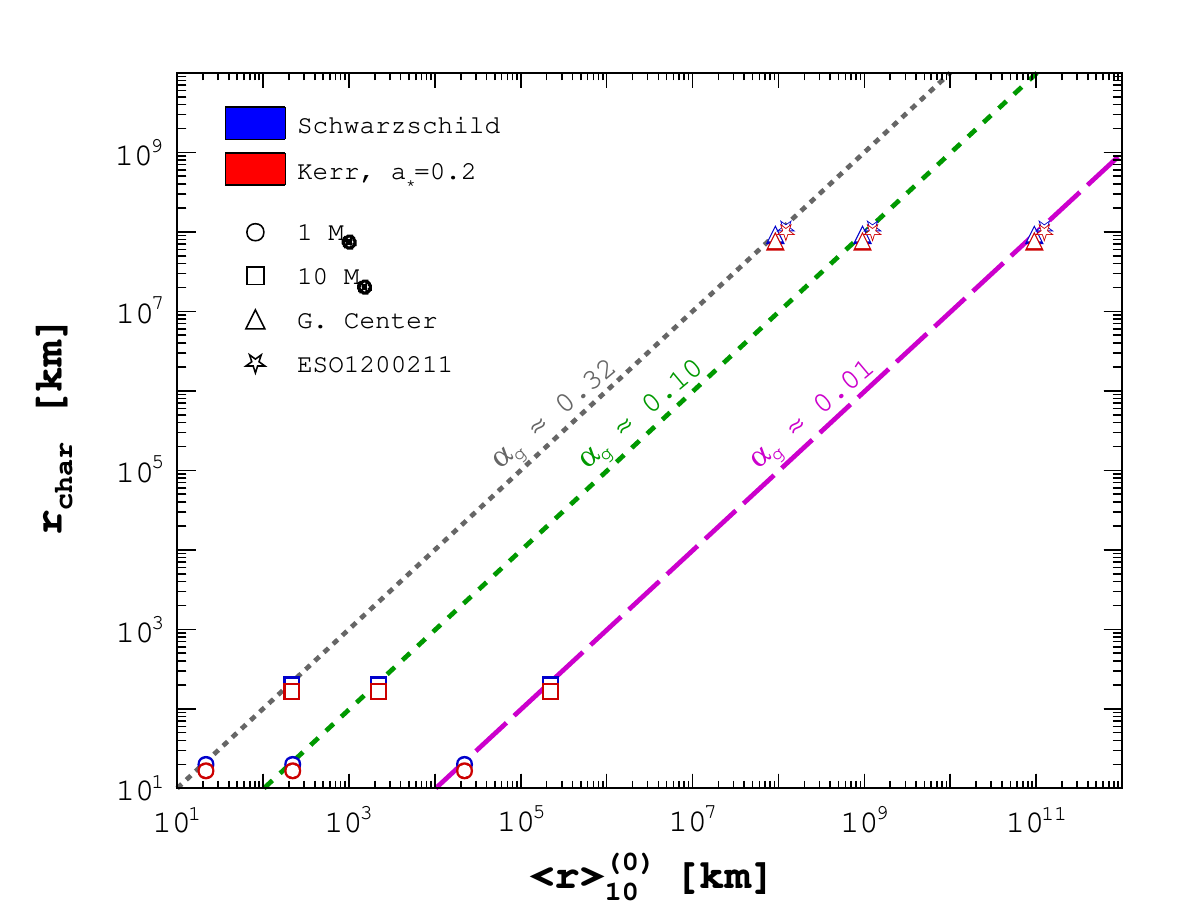}
    \end{minipage}
    \\[2 mm]
    \begin{minipage}{0.49\textwidth}
        \includegraphics[width=1.0\columnwidth]{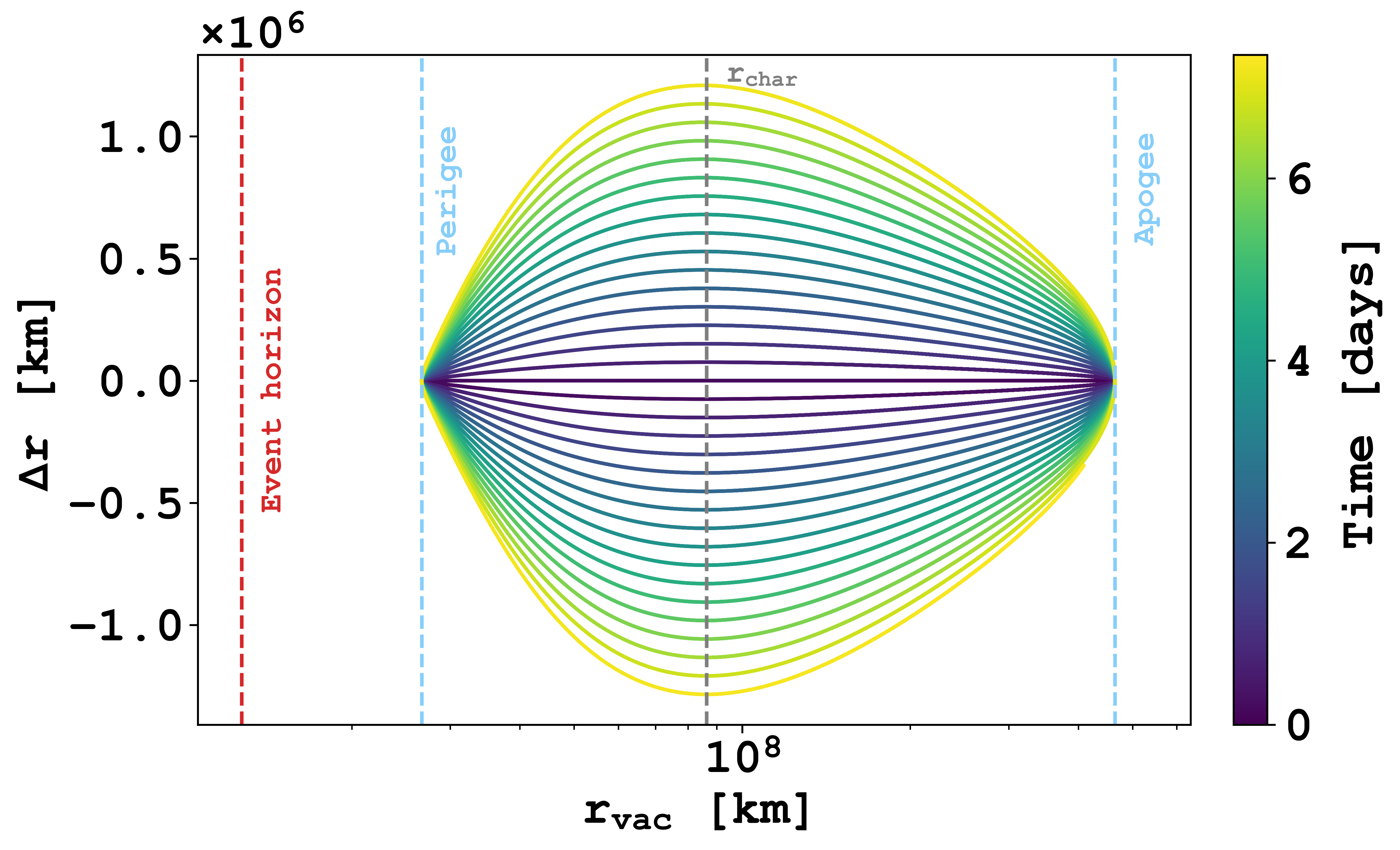}
    \\[2 mm]
        \includegraphics[width=1.0\columnwidth]{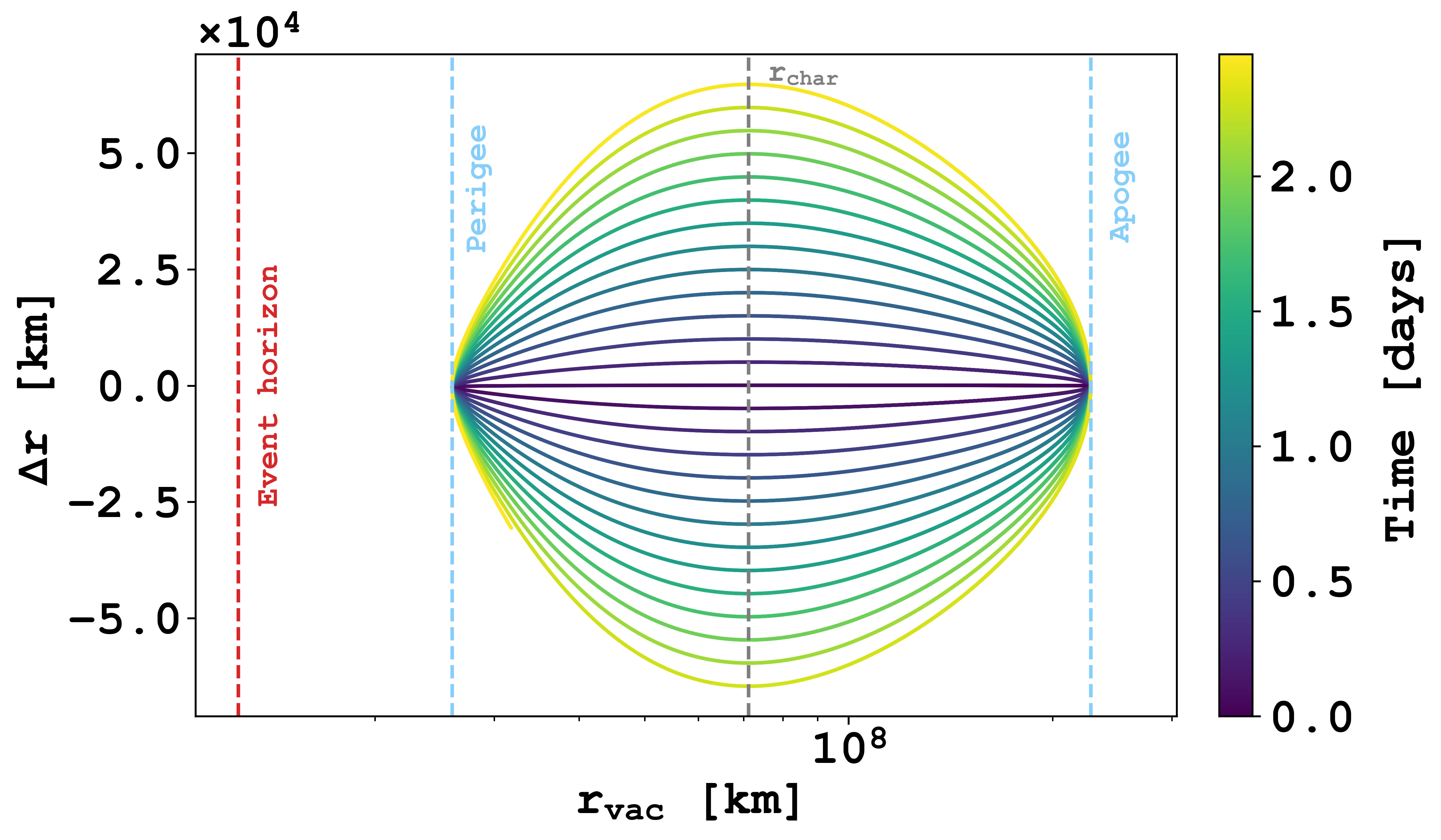}
    \end{minipage}
     \caption{
        Orbital response of gravitationally bound axion clouds.
        \textit{Top:} Characteristic response radius $r_{\rm char}$ versus the hydrogenic ground-state radius $\langle r\rangle_{10}^{(0)}$ for the black hole benchmarks considered. Marker shapes identify the source, while blue and red denote Schwarzschild and Kerr ($a_\ast=0.2$), respectively. Dashed lines indicate fixed gravitational couplings $\alpha_g\simeq0.32$, $0.10$, and $0.01$, with the markers showing the corresponding numerical geodesic results. The persistence of $r_{\rm char}\simeq\langle r\rangle_{10}^{\rm SFDM}$ toward smaller $\alpha_g$ demonstrates the robustness of the correspondence as the hydrogenic approximation becomes increasingly well controlled.
        \textit{Middle:} Radial deviation $\Delta r$ from the corresponding
        vacuum orbit for the Galactic-center Schwarzschild benchmark.
        \textit{Bottom:} Same for Kerr with $a_\ast=0.2$.
        The vertical dashed line marks $r_{\rm char}$, while the cyan lines indicate the orbital perigee and apogee.
    }
\label{fig:radial-phi-sim}
\end{figure}

In order to quantify this behaviour we define
\begin{equation}
    \Delta r_{\mathrm{geod}}(\tau)
        \equiv
        r_{\rm vac}(\tau)
        -
        r_{\rm SFDM}(\tau)
        \, ,
\end{equation}
and identify the characteristic orbital-response radius as
\begin{equation}
\begin{aligned}
    r_{\rm char}
        &=
        r_{\rm vac}(\tau_{\rm max})
        \, ,
    \\[2mm]
    \tau_{\rm max}
        &=
        \operatorname*{arg\,max}_{\tau}
        \Delta r_{\mathrm{geod}}(\tau)
        \, .
\end{aligned}
\label{eq:rchar}
\end{equation}
namely the radius along the reference vacuum trajectory at which the
orbital deviation is maximal (see Fig.~\ref{fig:radial-phi-sim}
and the {\it Supplemental Material} for representative geodesic
simulations).

The characteristic radius $r_{\rm char}$ lies between the perigee
and apogee of the reference orbit and represents an orbital-response
scale rather than an intrinsic scale of the DM distribution. For
multi-leaf trajectories, $r_{\rm char}$ is consistently recovered over
successive $4\pi$ cycles, corresponding to complete traversals of the
orbital pattern.

The perturbative analysis predicts that the orbital response should be
governed by the perturbed spatial extent of the gravitational atom. We
therefore test whether the numerically determined response radius
satisfies
\begin{equation}
    r_{\rm char}
        \simeq
        \langle r\rangle_{n\ell}^{\rm SFDM}
        \, ,
    \label{eq:prediction}
\end{equation}
where $\langle r\rangle_{n\ell}^{\rm SFDM}$ is obtained independently
from first-order perturbation theory.

This correspondence has a simple physical origin: the perturbation
redistributes the cloud density, localizing the change in enclosed
gravitational mass over the same radial region that controls
$\langle r\rangle_{n\ell}^{\rm SFDM}$ (see Supplemental Material).

The agreement is remarkable because the perturbative and numerical
approaches probe independent aspects of the system. The former predicts
the spatial extent of the perturbed gravitational atom, whereas the
latter determines the characteristic radius directly from classical
geodesic motion. For scenarios with equivalent $\alpha_g$, this
agreement is strictly linear with unit slope, demonstrating a one-to-one
correspondence between $r_{\rm char}$ and $\langle r\rangle_{n\ell}^{\rm SFDM}$. Their agreement demonstrates that the orbital-response
radius tracks the perturbed axion-cloud radius, revealing a purely
gravitational signature of scalar-field dark matter around primordial
black holes.

\vspace{0.2cm}

\noindent\textit{Observable Signatures\,---\,}Although the present analysis is restricted to test-particle dynamics, the perturbative redistribution of gravitationally bound axion clouds naturally gives rise to observable consequences. For an occupied cloud of mass $M_{\rm cl}$ and normalised wavefunction $\psi_{n\ell m}$, the density $\rho_{\rm cl}$ and density perturbation $\delta \rho_{\rm cl}$ are
\begin{equation}
    \rho_{\rm cl}
        =
            M_{\rm cl}\big| \psi_{n\ell m} \big|^2
            \, ,
    \quad
    \delta\rho_{\rm cl}
        =
            2\,M_{\rm cl}\,
            {\rm Re}
            \!\left[
                \psi^{(0)*}_{n\ell m}\,
                \delta\psi_{n\ell m}
            \right]
            ,
\end{equation}
which conserves the total cloud mass,
\begin{equation}
    \int {\rm d}^{3}x\,\delta\rho_{\rm cl}
        =
            0
            \, ,
\end{equation}
while redistributing its spatial profile. The corresponding perturbation of the enclosed mass,
\begin{equation}
    \delta M_{\rm cl}(<r)
        =
            4\pi
            \int_{0}^{r}
            {\rm d}r'\,
            r'^2\.\.
            \delta\rho_{\rm cl}(r')
            \, ,
\end{equation}
modifies invariant orbital observables such as the fundamental frequencies, periastron advance, and gravitational redshift.

A particularly promising application is to binaries containing black holes dressed by gravitational atoms. The redistribution of the cloud changes the gravitational potential sampled by an inspiralling companion and may therefore imprint characteristic signatures on the gravitational-wave phase evolution. Similar studies have already been carried out for particle dark-matter spikes surrounding black holes~\cite{Kavanagh:2020, Coogan:2022}; extending them to gravitational atoms embedded in scalar-field dark matter provides a direct avenue for testing the framework developed here. For isolated primordial black holes, microlensing offers a complementary probe whenever the cloud size approaches the Einstein radius. The redistribution also affects axion--photon conversion, governed by the interaction
\begin{equation}
    {\cal L}_{a\gamma}
        =
            -\frac{g_{{\rm a}\gamma\gamma}}{4}\,
            a F_{\mu\nu}\widetilde{F}^{\mu\nu}
            \, ,
\end{equation}
where $g_{{\rm a}\gamma\gamma}$ is the axion--photon coupling, $F_{\mu\nu}=\partial_\mu A_\nu-\partial_\nu A_\mu$ is the electromagnetic field-strength tensor, and $\widetilde{F}^{\mu\nu}=\frac{1}{2}\epsilon^{\mu\nu\rho\sigma}F_{\rho\sigma}$ is its dual. The emitted luminosity consequently depends on the overlap between the axion-cloud density and the resonant conversion region.
\begin{equation}
    L_{\gamma}
        \propto
            g_{{\rm a}\gamma\gamma}^2
            \int {\rm d}^{3}x\,
            B_{\perp}^2\,
            \rho_{\rm c}\,
            {\cal W}_{\rm pl}
            \, ,
\end{equation}
where ${\cal W}_{\rm pl}$ is the plasma-response function, determined by the local plasma frequency $\omega_{\rm p}$ together with coherence, absorption, and photon propagation effects. The perturbative redistribution derived here therefore enhances or suppresses the observable signal depending on the magnetic-field and plasma configuration~\cite{Ferreira:2024}. For ultralight axions ($m_{\rm a} \lesssim 10^{-12}\,\mathrm{eV}$), the corresponding photon frequencies lie below the terrestrial ionospheric cutoff, making oscillatory birefringence a more promising observational probe than conventional radio searches. A quantitative assessment of these signatures, including the dynamical evolution of the occupied cloud and its gravitational backreaction, is left for future work.

\vspace{0.2cm}
\noindent\textit{Conclusions\,---\,}We have shown that the gravitational response of axion dark matter to primordial black holes is governed by the perturbative redistribution of gravitationally bound states. Using first-order perturbation theory, we demonstrated that the surrounding scalar-field halo modifies the spatial structure of the gravitational atom, while numerical integrations of timelike geodesics reveal that the maximum orbital response consistently occurs at the perturbed cloud radius.

A central result of this work is that the characteristic orbital-response radius tracks the perturbed expectation value of the gravitational atom. The agreement between the perturbative prediction and the geodesic simulations establishes a direct correspondence between the microscopic structure of gravitationally bound axion clouds and the macroscopic dynamics of orbiting particles. Rather than introducing a new length scale, the scalar-field halo imprints its structure directly onto classical orbital motion through gravity alone, independently of the non-gravitational couplings of the axion.

The perturbative framework developed here can readily be extended to other dark-matter distributions, including Navarro--Frenk--White halos, as illustrated in the Supplemental Material. More generally, the correspondence identified in this work opens a new avenue for probing ultralight dark matter through the gravitational environments of compact objects.

\vspace{0.1cm}
\noindent\textit{Acknowledgments.---}J.~Ruz, J.~K.~Vogel and E.~Stamou acknowledge support from the Deutsche Forschungsgemeinschaft (DFG, German Research Foundation) under Germany's Excellence Strategy -- Cluster of Excellence “Color meets Flavor”, EXC 3107 -- Project-ID 533766364. J.~K.~Vogel also acknowledges funding by the German federal and state program ``Professorinnenprogramm 2030" Project-ID 01FP24167Q. C.~K{{\"o}}hn is funded by the Independent Research Fund Denmark (grant 1054-00104) and is funded by the Deutsche Forschungsgemeinschaft (DFG, German Research Foundation) - 548340212. V.~Mungai acknowledges support from Erasmus Mundus Joint Master (EMJM) Scholarship from the European Commission.

\bibliographystyle{apsrev4-2}
\bibliography{mybib_arXiv_clean}

\clearpage

\setcounter{page}{1}
\setcounter{figure}{0}
\setcounter{equation}{0}
\setcounter{table}{0}
\renewcommand{\theequation}{S\arabic{equation}}
\renewcommand{\thepage}{S\arabic{page}}
\renewcommand{\thefigure}{\Alph{figure}}

\clearpage
\onecolumngrid
\begin{center}
  \large\textbf{Supplemental Material}
\end{center}
\vspace{0.6cm}

\section{Perturbative Corrections to Hydrogenic Axion Bound States in Scalar-Field Dark Matter Halo Spacetimes}

\subsection{Klein--Gordon Equation in the SFDM Background}
\label{SM:KG}

In this Supplemental Material we derive the leading-order corrections to hydrogenic axion bound states induced by the gravitational field of a scalar-field dark matter (SFDM) halo. We begin from the covariant Klein--Gordon equation for a massive scalar field propagating on the background spacetime generated by a black hole embedded in a SFDM halo.

The dynamics of the scalar axion field $\Phi$ are governed by
\begin{equation}
	\left(\Box-m_{\rm a}^{2}\right)\Phi
		=
			0\, ,
	\label{eq:KG_SM}
\end{equation}
where $m_{\rm a}$ denotes the axion mass, and
\begin{equation}
	\Box
		\equiv
			\frac{1}{\sqrt{-g}}\,\partial_{\mu}\!\left(\sqrt{-g}\, g^{\mu\nu} \partial_{\nu} \right)
\end{equation}
is the covariant d'Alembert operator associated with the background metric $g_{\mu\nu}$. Throughout this work we consider the static, spherically symmetric metric
\begin{equation}
	{\rm d}s^{2}
		=
			-f(r)\,{\rm d}t^{2} + f(r)^{-1}\,{\rm d}r^{2} + r^{2}\!\left({\rm d}\theta^{2} + \sin^{2}\theta\,{\rm d}\phi^{2}\right),
	\label{eq:metric_SM}
\end{equation}
with
\begin{equation}
	f(r)
		=
			\exp[-\Phi_h(r)] - \frac{2M_{\mathrm{BH}}}{r}\, ,
	\label{eq:fmetric_SM}
\end{equation}
where $M_{\mathrm{BH}}$ denotes the black-hole mass, and
\begin{equation}
	\Phi_h(r)
		=
			\frac{8\rho_{\rm c}\.\.R^{2}}{\pi} \frac{\sin(\pi\.\.r/R)}{\pi\.\.r/R}
	\label{eq:halo_profile}
\end{equation}
characterises the gravitational contribution of the scalar-field dark-matter halo. Here $\rho_{\rm c}$ is the central halo density and $R$ denotes the condensate radius. Since the metric determinant is $g=-r^{4}\sin^{2}\theta$, we have $\sqrt{-g}=r^{2}\sin\theta$. Substituting Eq.~\eqref{eq:metric_SM} into the covariant d'Alembertian, the Klein--Gordon equation for the black-hole--dark-matter-halo system becomes
\begin{equation}
	-\frac{1}{f(r)} \frac{\partial^{2}\Phi}{\partial t^{2}} + \frac{1}{r^{2}} \frac{\partial}{\partial r}\!\left[r^{2}f(r)\,\frac{\partial\Phi}{\partial r}\right] + \frac{1}{r^{2}\sin\theta}\frac{\partial}{\partial\theta} \left(\sin\theta \frac{\partial\Phi}{\partial\theta}\right) + \frac{1}{r^{2}\sin^{2}\theta}\, \frac{\partial^{2}\Phi}{\partial\phi^{2}} - m_{\rm a}^{2}\Phi
		=
			0\, .
	\label{eq:KG_expanded}
\end{equation}

Equation~\eqref{eq:KG_expanded} is the complete equation for the static SFDM spacetime described by Eq.~\eqref{eq:metric_SM}. To solve it, we assume that the spacetime remains stationary and spherically symmetric. Under these conditions, the scalar field admits the separable ansatz
\begin{equation}
	\Phi(t,r,\theta,\.\.\phi)
		=
			e^{-i\omega t}\,Y_{\ell m}(\theta,\.\.\phi) R_{n\ell}(r)\, ,
	\label{eq:ansatz_SM}
\end{equation}
where $Y_{\ell m}$ are the usual spherical harmonics satisfying
\begin{equation}
	\nabla^{2}_{\Omega} Y_{\ell m}
		=
			-\ell(\ell+1) Y_{\ell m}\, .
\end{equation}

The parameter $\omega$ is the separation constant associated with the time dependence of the scalar field. Physically, it represents the oscillation frequency (or energy eigenvalue) of the bound-state solution. Imposing the appropriate boundary conditions at the black-hole horizon and at spatial infinity restricts $\omega$ to a discrete spectrum of allowed eigenfrequencies. The insertion of Eq.~\eqref{eq:ansatz_SM} into Eq.~\eqref{eq:KG_expanded} yields the radial equation
\begin{equation}
	\frac{1}{r^{2}} \frac{{\rm d}}{{\rm d}r}\! \left(r^{2}f(r)\,\frac{{\rm d}R_{n\ell}}{{\rm d}r}\right) + \left[\frac{\omega^{2}}{f(r)} - m_{\rm a}^{2} - \frac{\ell(\ell+1)}{r^{2}}\right] R_{n\ell}
		=
			0\, ,
	\label{eq:radial_exact}
\end{equation}
that constitutes the exact bound-state equation for a massive scalar field in the Schwarzschild black-hole spacetime modified by the presence of a scalar-field dark matter halo.

At present, no analytic solutions are known for the general form of Eq.~\eqref{eq:radial_exact}. In the following section we therefore consider the weak-field limit appropriate for dilute SFDM halos, where the halo contribution can be treated perturbatively and the equation reduces to an effective Schrödinger problem.

\subsection{Weak-Field Limit and Effective Gravitational Potential}
\label{SM:weakfield}

The exact radial equation derived in the previous section cannot, in general, be solved analytically owing to the non-trivial radial dependence of the metric function $f(r)$. To obtain analytic insight, we consider the weak-field regime appropriate for dilute scalar-field dark-matter halos, where the halo contribution represents only a small perturbation of the vacuum Schwarzschild geometry.

For realistic halo densities, the dimensionless gravitational potential satisfies $\Phi_h(r)\ll1$ throughout the region of interest. Therefore, the exponential function in Eq.~\eqref{eq:fmetric_SM} may be expanded in a Taylor series about $\Phi_h=0$ as
\begin{equation}
	\exp[-\Phi_h(r)]
		=
			1 - \Phi_h(r) + \frac{\Phi_h^{2}(r)}{2} +\mathcal{O}\big( \Phi_h^{3}(r) \big)
            \, .
\end{equation}
Retaining only first-order terms yields
\begin{equation}
	f(r)
		\simeq
			1 - \Phi_h(r) - \frac{2M_{\mathrm{BH}}}{r}\, .
	\label{eq:f_linear}
\end{equation}

In the weak-field regime, we write
\begin{equation}
	g_{tt}
		=
			-f(r)
		\simeq
			-\big( 1 + 2\Psi(r) \big)
            \, ,
	\label{eq:newton_metric}
\end{equation}
where $\Psi(r)$ denotes the Newtonian gravitational potential. Since $g_{tt}=-f(r)$, Eq.~\eqref{eq:f_linear} implies that
\begin{equation}
	g_{tt}
		\simeq
			-\left[1 - \frac{2M_{\mathrm{BH}}}{r} - \Phi_h(r)\right].
\end{equation}

The total Newtonian potential therefore becomes $\Psi(r) = \Psi_{\mathrm{BH}}(r) + \Psi_{\mathrm{SFDM}}(r)$, providing the starting point for the non-relativistic reduction of the Klein--Gordon equation carried out in the next section. In our approximation, the scalar-field halo acts as an additional gravitational potential superimposed on the Schwarzschild background, allowing the bound-state problem to be treated within time-independent perturbation theory.

\subsection{Non-Relativistic Reduction and Effective Hamiltonian}
\label{SM:NR}

As discussed above, no closed-form analytic solutions are known for Eq.~\eqref{eq:radial_exact}. Nevertheless, in the regime relevant for dilute scalar-field dark-matter halos and gravitationally bound axions, the problem admits a systematic non-relativistic expansion. Throughout this section we assume
\begin{equation}
	\lvert\Psi(r)\rvert
		\ll
			1\, ,
	\qquad \alpha_{\rm g}
		\ll
			1\, ,
\end{equation}
so that both the gravitational field is weak and the gravitationally bound axion is non-relativistic. In the hydrogenic regime, the characteristic velocity of the bound state scales as $v\sim\alpha_{\rm g}$, such that $\alpha_{\rm g}\ll1$ implies $v\ll1$.
In this regime the metric function satisfies
\begin{equation}
	f(r)
		\simeq
			1+2\Psi(r)\, ,
\end{equation}
and the spacetime line element becomes
\begin{equation}
	{\rm d}s^{2}
		=
			-
            \big[
                1 + 2\Psi(r)
            \big]\.
            {\rm d}t^{2}
            +
            \big[
                1 + 2\Psi(r)
            \big]^{-1}
            \.{\rm d}r^{2} + r^{2}\,{\rm d}\Omega^{2}
            \, .
\end{equation}

Expanding the radial component to first order in the gravitational potential yields
\begin{equation}
	{\rm d}s^{2}
		\simeq
        -\big[ 1+2\Psi(r) \big]\,{\rm d}t^{2} + \big[ 1-2\Psi(r) \big]\,{\rm d}r^{2} + r^{2}\,{\rm d}\Omega^{2}\, ,
	\label{eq:metric_NR}
\end{equation}
which provides the starting point for the non-relativistic reduction of the Klein--Gordon equation.

Following the standard Foldy--Wouthuysen/Feshbach--Villars reduction of the Klein--Gordon equation \cite{Feshbach:1958,Ruffini:1969qy}, we separate the rapidly oscillating rest-mass contribution from the slowly varying envelope according to

\begin{equation}
    \Phi(t,\bm{x})
    =
    \frac{1}{\sqrt{2m_{\rm a}}}\,
    \psi(t,\bm{x})e^{-i m_{\rm a} t}
    \,+ \mathrm{h.c.}
    ,
    \label{eq:NRansatz}
\end{equation}
where the complex wavefunction $\psi$ satisfies
\begin{equation}
	\lvert\partial_t\psi\rvert
		\ll
			m_{\rm a}\lvert\psi\rvert\, ,
	\qquad \lvert\nabla\psi\rvert
		\ll
			m_{\rm a}\lvert\psi\rvert\, .
	\label{eq:slow}
\end{equation}

These inequalities express that the residual energy and momentum of the bound state are both much smaller than the axion rest-mass scale, consistent with the non-relativistic regime. The temporal derivatives of the scalar field then become
\begin{align}
	\partial_t\Phi(t,\bm{x})
		&=
			\frac{1}{\sqrt{2\.m_{\rm a}}}\,e^{-i\.\.m_{\rm a}t}\.
            \big[
                \partial_t\psi(t,\bm{x}) - i\.\.m_{\rm a}\.\.\psi(t,\bm{x})
            \big]
            \, ,
            \\[2mm]
	\partial_t^{2}\Phi(t,\bm{x})
		&=
			\frac{1}{\sqrt{2\.m_{\rm a}}}\,e^{-i\,m_{\rm a} t}
            \Big[
                \partial_t^{2}\psi(t,\bm{x}) - 2\,i\,m_{\rm a}\.\.\partial_t\psi(t,\bm{x}) - m_{\rm a}^{2}\.\.\psi(t,\bm{x})
            \Big]
            \, .
\end{align}
Since the envelope varies on time scales much longer than the Compton time, $\lvert\partial_t^{2}\psi\rvert \ll m_{\rm a}\lvert\partial_t\psi\rvert$, the second-order time derivative may consistently be neglected. Likewise, expanding the metric only to first order in the gravitational
potential,
\begin{equation}
	\lvert\Psi(r)\rvert
		\ll
			1\, ,
\end{equation}
the Klein--Gordon equation reduces to the Schrödinger equation describing a non-relativistic particle moving in the external gravitational potential $\Psi(r)$:
\begin{equation}
	i \frac{\partial\psi}{\partial t}
		=
			\left[-\frac{\nabla^{2}}{2\.m_{\rm a}} +m_{\rm a}\Psi(r)\right] \psi\, .
	\label{eq:Schrodinger_supp}
\end{equation}

Substituting the decomposition of the potential,
\begin{equation}
	\Psi(r)
		=
			-\frac{M_{\mathrm{BH}}}{r} + \Psi_{\mathrm{SFDM}}(r)\, ,
\end{equation}
gives
\begin{equation}
	i \frac{\partial\psi}{\partial t}
		=
			\left[-\frac{\nabla^{2}}{2\.m_{\rm a}} - \frac{M_{\mathrm{BH}}\.\.m_{\rm a}}{r} + m_{\rm a}\Psi_{\mathrm{SFDM}}(r)\right] \psi\, .
	\label{eq:Schrodinger_split}
\end{equation}

The effective Hamiltonian therefore separates naturally into
\begin{equation}
	H
		=
			H_{0} + \delta H_{\mathrm{SFDM}}\, ,
	\label{eq:Hsplit}
\end{equation}
where
\begin{equation}
	H_{0}
		=
			-\frac{\nabla^{2}}{2\.m_{\rm a}} - \frac{M_{\mathrm{BH}}\.\.m_{\rm a}}{r}\, ,
	\label{eq:H0_supp}
\end{equation}
is the Hamiltonian of the gravitational hydrogen atom, while
\begin{equation}
	\delta H_{\mathrm{SFDM}}
		=
			m_{\rm a}\Psi_{\mathrm{SFDM}}(r)\, ,
	\label{eq:dH1}
\end{equation}
represents the perturbation induced by the scalar-field halo. Using Eq.~\eqref{eq:Psi_halo}, the perturbation assumes the explicit form
\begin{equation}
	\delta H_{\mathrm{SFDM}}(r)
		=
			-\frac{4\.m_{\rm a}\rho_{\rm c}R^{2}}{\pi} \frac{\sin(\pi\.\.r/R)}{\pi\.\.r/R}\, .
	\label{eq:dH_SFDM}
\end{equation}

In the weak-field regime, the macroscopic gravitational field generated by the SFDM halo enters the axion bound-state problem as a spherically symmetric perturbation of the vacuum hydrogenic Hamiltonian. Unlike the self-gravity of the bound axion cloud considered in studies of boson stars or superradiance, the perturbation derived here originates from an external gravitational environment. Consequently, the problem may be treated using the standard machinery of time-independent perturbation theory. In the following section we compute the resulting shifts of the hydrogenic energy spectrum and the corresponding corrections to the characteristic size of the axion cloud. At this order we neglect terms involving gradients of the weak potential, which are higher order in the non-relativistic expansion.

\subsection{First-Order Perturbation Theory}
\label{SM:perturbation}

We now treat the SFDM contribution in Eq.~\eqref{eq:dH_SFDM} as a small
perturbation of the vacuum hydrogenic problem. The unperturbed Hamiltonian
is
\begin{equation}
	H_{0}
		=
			-\frac{\nabla^{2}}{2\.m_{\rm a}} - \frac{ M_{\mathrm{BH}}\.\.m_{\rm a} }{ r }
            \, ,
\end{equation}
with normalised eigenstates
\begin{equation}
	\psi^{(0)}_{n\ell m}(\bm{x})
		=
			R_{n\ell}^{(0)}(r)\.\.Y_{\ell m}(\theta,\.\.\phi)\, .
\end{equation}

Here, \(R_{n\ell}^{(0)}(r)\) denotes the radial wavefunction of the unperturbed
hydrogenic state with principal quantum number \(n\) and angular momentum
\(\ell\). Explicitly,
\begin{equation}
	R_{n\ell}^{(0)}(r)
		=
        \mathcal{C}_{n\ell} \left(\frac{2r}{n\.\.a_{\rm g}}\right)^{\!\ell}\.\. e^{-\frac{r}{n\.\.a_{\rm g}}}\.\.L_{n-\ell-1}^{2\ell+1} \left(\frac{2r}{n\.\.a_{\rm g}}\.\.\right),
	\label{eq:Rnl_def}
\end{equation}
where \(L_p^q\) is an associated Laguerre polynomial, \(\mathcal{C}_{n\ell}\) is the normalisation constant given by
\begin{equation}
	\mathcal{C}_{n\ell}
		=
			\left(\frac{2}{n \.\.a_{\rm g}}\right)^{\!3/2} \sqrt{\frac{(n-\ell-1)!}{2n\,(n+\ell)!}}\, ,
	\label{eq:Cnl}
\end{equation}
and
\begin{equation}
	a_{\rm g}
		=
			\frac{1}{\alpha_{\rm g}\.\.m_{\rm a}}
		=
			\frac{ 1 }{ M_{\mathrm{BH}}\.\.m_{\rm a}^{2} }
\end{equation}
is the gravitational Bohr radius. Throughout this work, we adopt the normalisation convention
\begin{equation}
	\int_{0}^{\infty} {\rm d}r\,r^{2}\lvert R_{n\ell}^{(0)}(r)\rvert^{2}
		=
			1\, ,
\end{equation}
and the corresponding unperturbed energy eigenvalues are
\begin{equation}
	E^{(0)}_n
		=
			-\frac{ m_{\rm a}\.\.\alpha_{\rm g}^{2} }{ 2n^{2} }
            \, .
	\label{eq:E0}
\end{equation}

For the purely Newtonian gravitational potential, $V(r) = m_{\rm a}\Psi_{\mathrm{BH}}(r) = -\alpha_{\rm g}/r$, the unperturbed Hamiltonian exhibits the same hidden symmetry as the Coulomb problem. Consequently, the energy eigenvalues depend only on the principal quantum number $n$ and are independent of the orbital angular momentum quantum number $\ell$. The corresponding eigenstates are labelled by $\lvert n \ell m\rangle$, with $\ell = 0,\,1,\,\ldots,\,n-1$ and $m = -\ell,\,\ldots,\,\ell$. The degeneracy with respect to $\ell$ is lifted only by perturbations that break the potential's $1/r$ symmetry, such as relativistic corrections or other non-Keplerian interactions. Therefore, the first-order correction to the energy is
\begin{equation}
	\Delta E_{n\ell m}^{\mathrm{SFDM}}
		=
			\big\langle n\ell m \big\rvert \delta H_{\mathrm{SFDM}} \big\lvert n\ell m\big\rangle\, .
	\label{eq:DE_def}
\end{equation}

Since the SFDM perturbation is spherically symmetric, it does not depend on $m$ and does not mix states with different $\ell$ or $m$. Hence, $\Delta E_{n\ell m} \equiv \Delta E_{n\ell}$. Using Eq.~\eqref{eq:dH_SFDM}, we obtain
\begin{equation}
	\Delta E_{n\ell}^{\mathrm{SFDM}}
		=
			-\frac{4\.m_{\rm a}\rho_{\rm c}R^{2}}{\pi} \int {\rm d}^{3}x\, \lvert R_{n\ell}(r)\rvert^{2}\,\big\lvert Y_{\ell m}(\theta,\.\.\phi)\big\rvert^{2}\,\frac{\sin(\pi\.\.r/R)}{\pi\.\.r/R}\, ,
\end{equation}
and, since the spherical harmonics satisfy $\int {\rm d}\Omega\,\lvert Y_{\ell m}\rvert^{2} = 1$, the energy shift reduces to
\begin{equation}
	\Delta E_{n\ell}^{\mathrm{SFDM}}
		=
			-\frac{4\.m_{\rm a}\rho_{\rm c}R^{2}}{\pi} \int_{0}^{\infty} {\rm d}r\, r^{2} \lvert R_{n\ell}(r)\rvert^{2} \frac{\sin(\pi\.\.r/R)}{\pi\.\.r/R}\, .
	\label{eq:DE_radial}
\end{equation}

Equation~\eqref{eq:DE_radial} gives the leading SFDM correction to the hydrogenic-like spectrum. The correction is negative when the bound state is localised mainly within the first positive lobe of the BEC profile, reflecting the additional attractive gravitational potential generated by the scalar halo. In addition, the first-order correction to the eigenstate is
\begin{equation}
	\lvert\delta\psi_{n\ell}\rangle
		=
			\sum_{n'\.\neq\.\.n} \frac{\big\langle n'\ell\big\rvert \delta H_{\mathrm{SFDM}} \big\lvert n\ell \big\rangle}{E^{(0)}_n - E^{(0)}_{n'}}\,
            \big\lvert n'\ell\big\rangle\, .
	\label{eq:dpsi_supp}
\end{equation}

Since the SFDM perturbation is spherically symmetric, it commutes with the angular momentum operators. Consequently, the orbital angular momentum quantum number \(\ell\) is conserved, and only hydrogenic states with the same \(\ell\), but different principal quantum number \(n'\), contribute to the first-order correction to the wavefunction. The perturbation matrix is therefore block diagonal in \(\ell\), with each block acting only on the radial quantum number,
\begin{equation}
	\delta H_{\mathrm{SFDM}}
		\sim
			\begin{pmatrix}
				\boxed{\delta H^{\ell=0}} & 0 & 0 & \cdots \\
				0 & \boxed{\delta H^{\ell=1}} & 0 & \cdots \\
				0 & 0 & \boxed{\delta H^{\ell=2}} & \cdots \\
				\vdots & \vdots & \vdots & \ddots
			\end{pmatrix}\, ,
\end{equation}
where
\begin{equation}
	\delta H^{\ell}
		=
			\begin{pmatrix}
				\delta E_{1}^{\ell} & \delta H_{12}^{\ell} & \delta H_{13}^{\ell} & \cdots \\
				\delta H_{21}^{\ell} & \delta E_{2}^{\ell} & \delta H_{23}^{\ell} & \cdots \\
				\delta H_{31}^{\ell} & \delta H_{32}^{\ell} & \delta E_{3}^{\ell} & \cdots \\
				\vdots & \vdots & \vdots & \ddots
			\end{pmatrix}\, .
\end{equation}
The radial matrix elements are
\begin{equation}
	\mathcal{V}^{\ell}_{n'n}
		=
			\big\langle n'\ell\big\rvert\delta H_{\mathrm{SFDM}}\big\lvert n\ell\big\rangle
		=
			-\frac{4\.m_{\rm a}\rho_{\rm c}R^{2}}{\pi} \int_{0}^{\infty} {\rm d}r\, r^{2} R_{n'\ell}(r) R_{n\ell}(r) \frac{\sin(\pi\.\.r/R)}{\pi\.\.r/R}\, ,
	\label{eq:Vmatrix}
\end{equation}
where the hydrogenic radial wavefunctions are given by Eq.~\eqref{eq:Rnl_def}.
The diagonal elements, \(\mathcal{V}^{\ell}_{nn}=\delta E_n^\ell\), yield the first-order energy shifts, whereas the off-diagonal elements (\(n'\.\neq\.\.n\)) mix states with identical angular momentum quantum numbers but different principal quantum number. Since \(R_{n\ell}(r)\) and \(R_{n'\ell}(r)\) have the same angular dependence but different radial nodes, radial extent, and unperturbed energies, the SFDM perturbation induces mixing only among these radial profiles, leading to the eigenstate correction given in Eq.~\eqref{eq:dpsi_supp}. Therefore,
\begin{equation}
	\lvert\delta\psi_{n\ell}\rangle
		=
			\sum_{n'\.\neq\.\.n} \frac{\mathcal{V}^{\ell}_{n'n}}{E^{(0)}_n-E^{(0)}_{n'}}\,\lvert n'\ell\rangle\, .
	\label{eq:dpsi_radial}
\end{equation}

This perturbation not only shifts the energy of the state, but it also redistributes the radial probability density of the axion cloud, whose expectation value becomes
\begin{equation}\label{eq:rad_add}
	\langle r\rangle_{n\ell}^{\mathrm{SFDM}}
		=
			\langle r\rangle_{n\ell}^{(0)} + \Delta\langle r\rangle_{n\ell}^{\mathrm{SFDM}}\, .
\end{equation}

Using the first-order corrected wavefunction,
\begin{equation}
	\big\lvert\tilde{\psi}_{n\ell}\big\rangle
		=
			\big\lvert n\ell\big\rangle + \big\lvert\delta\psi_{n\ell}\big\rangle + \mathcal{O}\big(\delta H_{\mathrm{SFDM}}^{2}\big)
            \, ,
\end{equation}
the expectation value of the orbital radius becomes
\begin{align}
	\langle r\rangle_{n\ell}^{\mathrm{SFDM}}
		&=
			\big\langle\tilde{\psi}_{n\ell}\big\rvert\,r\,\big\lvert\tilde{\psi}_{n\ell}\big\rangle \nonumber
            \\[2mm]
		&=
			\big\langle n\ell\big\rvert r\big\lvert n\ell\big\rangle +\big\langle\delta\psi_{n\ell}\big\rvert r\big\lvert n\ell\big\rangle +\big\langle n\ell\big\rvert r\big\lvert\delta\psi_{n\ell}\big\rangle +\mathcal{O}\big(\delta H_{\mathrm{SFDM}}^{2}\big)
            \, .
\end{align}
Retaining only terms linear in the perturbation therefore yields
\begin{equation}
	\Delta\big\langle r \big\rangle_{n\ell}^{\mathrm{SFDM}}
		=
			2\,\operatorname{Re} \big\langle n\ell\big\rvert r\big\lvert\delta\psi_{n\ell}\big\rangle
		=
			2\,\operatorname{Re} \sum_{n'\.\neq\.\.n} \frac{\mathcal{V}^{\ell}_{n'n}}{E^{(0)}_n-E^{(0)}_{n'}} \langle n\ell\rvert r\lvert n'\ell\rangle\, .
	\label{eq:Dr_general}
\end{equation}

Defining the radial overlap matrix
\begin{equation}
	\mathcal{R}^{\ell}_{nn'}
		=
			\int_{0}^{\infty} {\rm d}r\, r^{3}\, R_{n\ell}(r)\, R_{n'\ell}(r)\, ,
	\label{eq:Rmatrix}
\end{equation}
the first-order correction to the expectation value of the orbital radius can be written in the compact form
\begin{equation}
	\Delta\langle r\rangle_{n\ell}^{\mathrm{SFDM}}
		=
			2\,\operatorname{Re} \sum_{n'\.\neq\.\.n} \frac{\mathcal{V}^{\ell}_{n'n}\, \mathcal{R}^{\ell}_{nn'}}{E^{(0)}_n-E^{(0)}_{n'}}\, ,
	\label{eq:Dr_final}
\end{equation}
which shows that the radial displacement is governed by the interplay between the SFDM-induced mixing matrix $\mathcal{V}^{\ell}_{n'n}$ and the radial overlap matrix $\mathcal{R}^{\ell}_{nn'}$. Together with Eq.~\eqref{eq:rad_add}, this provides the perturbative prediction for the shift of the characteristic axion-cloud radius induced by the SFDM halo. In the main text, this prediction is compared with the radial region where the numerically integrated geodesics exhibit their largest deviations from the corresponding vacuum trajectories.

\subsection{Analytic Ground-State Correction for SFDM}
\label{SM:ground}

The perturbative formalism developed in the previous section may be illustrated explicitly for the lowest hydrogenic-like bound state, corresponding to the quantum numbers $n=1,\,\ell=0$. Besides providing a useful consistency check of the general formalism, the ground state admits a closed analytic expression for the first-order energy correction, allowing the dependence on the halo parameters to be made explicit. Taking into account that, in this particular case, the normalised radial wavefunction is
\begin{equation}
	R_{10}(r)
		=
			\frac{2}{a_{\rm g}^{3/2}}\,e^{-r/a_{\rm g}}\, ,
	\label{eq:R10}
\end{equation}
with the gravitational Bohr radius $a_{\rm g}=1/(M_{\mathrm{BH}}m_{\rm a}^{2})$. Substituting this definition into the first-order perturbation formula, Eq.~\eqref{eq:DE_radial}, yields
\begin{equation}
	\Delta E_{10}^{\mathrm{SFDM}}
		=
			-\frac{4\.m_{\rm a}\rho_{\rm c}R^{2}}{\pi} \int_{0}^{\infty} {\rm d}r\, r^{2} \lvert R_{10}(r)\rvert^{2}\,\frac{\sin(\pi\.\.r/R)}{\pi\.\.r/R}\, .
	\label{eq:DE10}
\end{equation}

Using the explicit form of the hydrogenic ground state gives
\begin{equation}
	\Delta E_{10}^{\mathrm{SFDM}}
		=
			-\frac{16m_{\rm a}\rho_{\rm c}R^{2}}{\pi a_{\rm g}^{3}} \int_{0}^{\infty} {\rm d}r\, r^{2} e^{-2r/a_{\rm g}}\, \frac{\sin(\pi\.\.r/R)}{\pi\.\.r/R}\, .
\end{equation}
Defining
\begin{equation}
	\beta
		\equiv
			\frac{2}{a_{\rm g}}\, ,
\end{equation}
the radial integral becomes
\begin{equation}
	\int_{0}^{\infty} {\rm d}r\, r^{2} e^{-\beta r} \frac{\sin(\pi\.\.r/R)}{\pi\.\.r/R}
		=
			\frac{R}{\pi} \int_{0}^{\infty} {\rm d}r\, r e^{-\beta r} \sin\!\left(\frac{\pi r}{R}\right)
		=
			\frac{2\beta}{\left[\beta^{2}+(\pi/R)^{2}\right]^{2}}\, .
\end{equation}
The resulting first-order energy shift is therefore
\begin{equation}
	\Delta E_{10}^{\mathrm{SFDM}}
		=
			-\frac{64\.m_{\rm a}\rho_{\rm c}R^{2}}{\pi a_{\rm g}^4} \frac{1}{\left[4/a_{\rm g}^{2}+\pi^{2}/R^{2}\right]^{2}}\, .
	\label{eq:DE10_final}
\end{equation}
The correction is always negative, reflecting the fact that the SFDM halo provides an additional attractive gravitational potential which increases the binding energy of the axion cloud.

Two limiting regimes follow naturally. For compact axion clouds ($a_{\rm g}\ll R$), the scalar field probes only the nearly homogeneous central region of the halo, giving
\begin{equation}
	\Delta E_{10}^{\mathrm{SFDM}}
		\simeq
			-\frac{4\.m_{\rm a}\rho_{\rm c}R^{2}}{\pi}\, .
\end{equation}
Conversely, when $a_{\rm g}\gtrsim R$, the cloud extends beyond the condensate radius and the perturbation becomes progressively suppressed owing to the oscillatory behaviour of the BEC density profile.

Although the energy correction derived above quantifies the strength of the halo-induced perturbation, it is not the quantity most directly relevant to the orbital dynamics discussed in the main text. We therefore derive the analytic correction to $\langle r\rangle$ in the following section.

\subsection{Analytic SFDM Correction to Mean Radius}
\label{SM:radius}

While the first-order energy shift depends only on the unperturbed ground-state wavefunction, the correction to the characteristic size of the cloud requires the first-order correction to the wavefunction itself. For the hydrogenic ground state $(n=1,\ell=0)$, the unperturbed expectation value of the radial coordinate is
\begin{equation}
	\langle r\rangle_{10}^{(0)}
		=
			\int {\rm d}^{3}x\, r\,\big\lvert\psi_{10}(\bm{x})\big\rvert^{2}
		=
			\int_{0}^{\infty} {\rm d}r\, r^{3} \big\lvert R_{10}(r)\big\rvert^{2}
		=
			\frac{3}{2}a_{\rm g}\, .
\end{equation}

To first order in the SFDM perturbation,
\begin{equation}
	\lvert\delta\psi_{10}\rangle
		=
			\sum_{n\neq1} \frac{\big\langle n0\big\rvert\delta H_{\mathrm{SFDM}}\big\lvert 10\big\rangle}{E^{(0)}_{1}-E^{(0)}_{n}} \big\lvert n0\big\rangle\, ,
\end{equation}
where only $\ell = 0$ intermediate states contribute because both the perturbing potential and the operator $r$ are spherically symmetric. The first-order correction to the expectation value therefore becomes
\begin{equation}
	\Delta\langle r\rangle_{10}^{\mathrm{SFDM}}
		=
			2 \sum_{n=2}^{\infty} \frac{\langle 10\rvert r\lvert n0\rangle \langle n0\rvert\delta H_{\mathrm{SFDM}}\lvert 10\rangle}{E^{(0)}_1-E^{(0)}_n}\, ,
	\label{eq:Dr_ground_sum}
\end{equation}
where
\begin{equation}
	\delta H_{\mathrm{SFDM}}
		\equiv
			\delta V(r)
		=
			-\frac{4\.m_{\rm a}\rho_{\rm c}R^{2}}{\pi} \frac{\sin(kr)}{kr}\, ,
	\qquad k
		=
			\frac{\pi}{R}\, .
\end{equation}
The perturbation matrix element is
\begin{equation}
	\big\langle n0\big\rvert\delta V(r)\big\lvert 10\big\rangle
		=
			-\frac{4\.m_{\rm a}\rho_{\rm c}R^{2}}{\pi} \int_{0}^{\infty} {\rm d}r\, r^{2} R_{n0}(r)\.\.R_{10}(r)\,\frac{\sin(kr)}{kr}\, ,
\end{equation}
while the radial matrix element is
\begin{equation}
	\big\langle 10\big\rvert r\big\lvert n0\big\rangle
		=
			\int_{0}^{\infty} {\rm d}r\, r^{3} R_{10}(r)\.\.R_{n0}(r)\, .
\end{equation}
Substituting these expressions into Eq.~\eqref{eq:Dr_ground_sum} yields
\begin{equation}
	\Delta\langle r\rangle_{10}^{\mathrm{SFDM}}
		=
			-\frac{8m_{\rm a}\rho_{\rm c}R^{2}}{\pi} \sum_{n=2}^{\infty} \frac{I_n(k)\, J_n}{E_{1}^{(0)}-E_n^{(0)}}\, ,
	\label{eq:Dr_ground_final}
\end{equation}
where
\begin{align}
	I_n(k)
		&=
			\int_{0}^{\infty} {\rm d}r\, r^{2} R_{n0}(r)\.\.R_{10}(r)\,\frac{\sin(kr)}{kr}\, ,
            \\[2mm]
	J_n
		&=
			\int_{0}^{\infty} {\rm d}r\, r^{3} R_{10}(r)\.\.R_{n0}(r)\, .
\end{align}

Equation~\eqref{eq:Dr_ground_final} constitutes the exact first-order perturbative prediction for the shift of the mean orbital radius. Unlike the energy correction, which depends only on the ground-state density, the radial correction depends on the first-order deformation of the wavefunction through its coupling to the complete hydrogenic spectrum. An equivalent representation is obtained by introducing the reduced Coulomb Green's function,
\begin{equation}
	G_{0}'(r,r';E_{1})
		=
			\sum_{n\neq1} \frac{R_{n0}(r)\.\. R_{n0}(r')}{E_{1}^{(0)}-E_n^{(0)}}\, ,
\end{equation}
which allows Eq.~\eqref{eq:Dr_ground_final} to be written as
\begin{equation}
	\Delta\langle r\rangle_{10}^{\mathrm{SFDM}}
		=
			2 \int_{0}^{\infty}{\rm d}r\, r^{3} R_{10}(r) \int_{0}^{\infty}{\rm d}r'\; {r'}^{2}\.\.G_{0}'(r,r';E_{1})\.\. \delta V(r')\.\.R_{10}(r')
            \, .
\end{equation}

This expression is exact and provides the natural starting point for either numerical evaluation or further analytic approximations.

Using the Dalgarno--Lewis method, the first-order correction to the expectation value of the radius can be obtained without explicitly summing over the complete hydrogenic spectrum. For a spherically symmetric perturbation, one finds
\begin{equation}
	\Delta\langle r\rangle_{10}^{\mathrm{SFDM}}
		=
			-m_{\rm a} a_{\rm g} \Big[\big\langle r^{2}\delta V(r)\big\rangle_{10} - \big\langle r^{2}\big\rangle_{10}\.\. \big\langle\delta V(r)\big\rangle_{10}\Big].
	\label{eq:Dr_DL}
\end{equation}

For the hydrogenic ground state,
\begin{equation}
	\langle r^{2}\rangle_{10}
		=
			3a_{\rm g}^{2}\, ,
\end{equation}
while the perturbing potential is
\begin{equation}
	\delta V(r)
		=
			-\frac{4\.m_{\rm a}\rho_{\rm c}R^{2}}{\pi} \frac{\sin(kr)}{kr}\, ,
	\qquad k
		=
			\frac{\pi}{R}\, .
\end{equation}
The required expectation values are
\begin{equation}
	\big\langle\delta V(r)\big\rangle_{10}
		=
			-\frac{64\.m_{\rm a}\rho_{\rm c}R^{2}}{\pi a_{\rm g}^4} \frac{1}{\left(4/a_{\rm g}^{2}+k^{2}\right)^{2}}\, ,
\end{equation}
and
\begin{equation}
	\big\langle r^{2}\delta V(r) \big\rangle_{10}
		=
			-\frac{768m_{\rm a}\rho_{\rm c}R^{2}}{\pi a_{\rm g}^4} \frac{4/a_{\rm g}^{2}-k^{2}}{\left(4/a_{\rm g}^{2}+k^{2}\right)^4}\, .
\end{equation}
Substituting these expressions into
Eq.~\eqref{eq:Dr_DL} yields
\begin{equation}
	\Delta\langle r\rangle_{10}^{\mathrm{SFDM}}
		=
			-192\pi\,m_{\rm a}^{2}\rho_{\rm c}\,a_{\rm g}^5 \frac{12+\pi^{2}\eta^{2}}{\left(4+\pi^{2}\eta^{2}\right)^4}\, ,
	\qquad \eta
		=
			\frac{a_{\rm g}}{R}
            \, .
	\label{eq:Dr10_final}
\end{equation}

In the compact-cloud limit, where $a_{\rm g}\ll R$, the result reduces to $\Delta\langle r\rangle_{10}\simeq -9\pi\,m_{\rm a}^{2}\rho_{\rm c}\,a_{\rm g}^5$. The negative sign indicates that the additional gravitational attraction of the SFDM halo contracts the ground-state wavefunction, reducing its
mean radius.

\subsection{Characteristic Scale of SFDM Condensate around Stellar-Mass Black Holes}
\label{sec:SFDM_solar_BH}

The equilibrium configuration of a scalar-field dark-matter (SFDM) condensate surrounding a black hole is, in general, determined by the interplay between the black hole gravitational field, the self-gravity of the condensate, quantum pressure, and possible scalar self-interactions. Accordingly, the characteristic size of the condensate cannot be inferred from the black hole mass alone, but also depends on the scalar mass, the total condensate mass, and the strength of the self-interaction. In the non-relativistic regime, these equilibrium configurations are described by the coupled Gross--Pitaevskii--Poisson (GPP) equations, which provide the standard theoretical framework for self-gravitating Bose--Einstein condensates and fuzzy-dark-matter halos \cite{Boehmer:2007um, Chavanis:2011zi, Hui2017, Marsh:2015xka}.

The analysis presented in the main text adopts a different, perturbative viewpoint. Rather than solving the GPP equations self-consistently, the SFDM halo is treated as an externally prescribed equilibrium background whose gravitational field produces a small correction to the black hole metric. The corresponding Newtonian potential, $\Psi_{\rm SFDM}(r)$, is therefore regarded as a fixed function when solving for the black hole-bound scalar state. Within this prescribed-background approximation, the Klein--Gordon equation reduces, in the weak-field and non-relativistic limits, to a linear Schr\"odinger equation with the halo contributing through the perturbing potential
\begin{equation}
	\delta V(r)
		=
			m_{\rm a}\Psi_{\rm SFDM}(r)\,.
	\label{eq:deltaV_halo_connection}
\end{equation}
This approach isolates the gravitational response induced by the halo while retaining the analytical simplicity of the hydrogenic black hole-bound problem.

The relationship between the present treatment and the fully self-consistent GPP description is well understood. Numerical studies of self-gravitating scalar condensates including a central point-mass potential have shown that the black hole compresses the central solitonic core and, in the regime where the black hole potential dominates over the self-gravity of the condensate, the equilibrium solution approaches the hydrogen-like bound states of a particle in a Newtonian \(1/r\) potential \cite{Davies:2019wgi}. The perturbative framework developed here therefore corresponds to the limiting case in which the background halo is treated as an externally prescribed source while the bound-state dynamics are governed primarily by the black hole potential. More recently, this connection has also been investigated for rotating and self-gravitating scalar configurations around black holes \cite{PalomaresChavez2024}.

\vspace{0.2 cm}
\paragraph{{\it \textbf{Black Hole-dominated, non-interacting regime:}}}

The perturbative framework developed in the main text corresponds to the regime in which the gravitational field of the black hole dominates the local dynamics of the bound scalar state, while the surrounding SFDM halo acts only as a weak perturbation. In this limit, both the self-gravity of the condensate and the scalar self-interaction are negligible compared to the Newtonian potential of the black hole, and the Gross--Pitaevskii equation therefore reduces to the unperturbed Hamiltonian employed throughout the main text,
\begin{equation}
	H_0
		=
			-\frac{\nabla^2}{2m_{\rm a}}-\frac{GM_{\rm BH}m_{\rm a}}{r}\,,
	\label{eq:H0_hydrogenic}
\end{equation}
whose stationary states satisfy the time-independent Schr\"odinger equation,
\begin{equation}
	\left[-\frac{\nabla^2}{2m_{\rm a}}-\frac{GM_{\rm BH}m_{\rm a}}{r}\right]\psi
		=
			E\psi\,.
	\label{eq:Schrodinger_BH}
\end{equation}

This eigenvalue problem is mathematically identical to the hydrogen atom, with the Coulomb interaction replaced by the Newtonian gravitational potential. Consequently, the bound-state spectrum is hydrogenic, with the electromagnetic fine-structure constant replaced by the gravitational coupling $\alpha_{g}$~\cite{Detweiler:1980, Arvanitaki:2010sy, Brito:2015oca}. The same hydrogenic limit also emerges from the self-consistent Schr\"odinger--Poisson equations describing fuzzy-dark-matter solitons when the gravitational field of the central black hole dominates over the self-gravity of the condensate \cite{Davies:2019wgi}.

The hydrogenic approximation is controlled by the condition $\alpha_{\rm g}\ll1$, which ensures that the bound-state velocity is non-relativistic and that the characteristic size of the cloud is much larger than the black hole horizon \cite{Detweiler:1980, Arvanitaki:2010sy, Brito:2015oca}. The hydrogenic ground state provides the reference configuration about which the perturbative treatment developed in the main text is constructed. Since the perturbation modifies primarily the spatial extent of the bound state, a natural measure of the cloud size is the expectation value of the orbital radius,
\begin{equation}
	R_{\rm c}
		\equiv
			\langle r \rangle_{10}\,,
\end{equation}
which serves as the characteristic radius throughout this work:
\begin{equation}
	R_{\rm c}
		=
			\langle r \rangle_{10}
		=
			\frac{3}{2}\,a_{\rm g}\,,
	\label{eq:Rc_ground}
\end{equation}
with corresponding mass-density profile
\begin{equation}
	\rho(r)
		=
			m_{\rm a}\big|\psi_{10}(r)\big|^2
		=
			\frac{M_{\rm c}}{\pi a_{\rm g}^3}\,e^{-2r/a_{\rm g}}\,,
	\label{eq:BH_bound_density}
\end{equation}
where $M_{\rm c} = N m_{\rm a}$ denotes the total condensate mass.

It is important to emphasize that \(R_{\rm c}\) characterizes the spatial extent of a black hole-bound scalar state. It should therefore not be confused with the kiloparsec-scale core radius commonly used to describe galactic SFDM halos \cite{Hui2017, Marsh:2015xka}. In the perturbative framework of the main text, the halo modifies this characteristic radius through the first-order correction $\Delta\langle r \rangle$, which provides the primary observable used to quantify the gravitational response of the bound state.

In order to illustrate the characteristic scales entering our numerical calculations, consider a solar-mass black hole,
\begin{equation}
	M_{\rm BH}
		=
			M_\odot\,,
	\qquad
	r_{\rm g}
		=
			1.477\,{\rm km}\,,
\end{equation}
the gravitational coupling becomes
\begin{equation}
	\alpha_{\rm g}
		\simeq
			0.748\left(\frac{M_{\rm BH}}{M_\odot}\right)\left(\frac{m_{\rm a}}{10^{-10}\ {\rm eV}}\right),
	\label{eq:alpha_solar}
\end{equation}
while the gravitational Bohr radius and the characteristic size of the ground state are
\begin{align}
	a_{\rm g}
		&\simeq
			2.64~{\rm km}\left(\frac{10^{-10}\ {\rm eV}}{m_{\rm a}}\right)^{\!2}\left(\frac{M_\odot}{M_{\rm BH}}\right),
	\\[2mm]
	R_{\rm c}
		&\simeq
			3.96~{\rm km}\left(\frac{10^{-10}\ {\rm eV}}{m_{\rm a}}\right)^{\!2}\left(\frac{M_\odot}{M_{\rm BH}}\right).
	\label{eq:Rc_numerical}
\end{align}
Representative values are summarised in Table~\ref{tab:solar_BH_cloud}, illustrating the rapid increase of the bound-state size with decreasing axion mass.

\begin{table}[!h]
\centering
{
\renewcommand{\arraystretch}{1.4}
\setlength{\tabcolsep}{14pt}
\begin{tabular}{cccc}
	\hline\hline
	\(m_{\rm a}\,[{\rm eV}]\)
		& \(\alpha_{\rm g}\)
		& \(a_{\rm g}\)
		& \(R_{\rm c}=3a_{\rm g}/2\)
		\\
	\hline
	\(10^{-10}\)
		& \(7.48\times10^{-1}\)
		& \(2.64~{\rm km}\)
		& \(3.96~{\rm km}\)
		\\
	\(10^{-11}\)
		& \(7.48\times10^{-2}\)
		& \(2.64\times10^{2}~{\rm km}\)
		& \(3.96\times10^{2}~{\rm km}\)
		\\
	\(10^{-12}\)
		& \(7.48\times10^{-3}\)
		& \(2.64\times10^{4}~{\rm km}\)
		& \(3.96\times10^{4}~{\rm km}\)
		\\
	\(10^{-13}\)
		& \(7.48\times10^{-4}\)
		& \(2.64\times10^{6}~{\rm km}\)
		& \(3.96\times10^{6}~{\rm km}\)
		\\
	\(10^{-14}\)
		& \(7.48\times10^{-5}\)
		& \(2.64\times10^{8}~{\rm km}\)
		& \(3.96\times10^{8}~{\rm km}\)
		\\
	\hline\hline
\end{tabular}}
\caption{Characteristic properties of the hydrogenic ground state surrounding a $1\,M_\odot$ black hole in the black hole-dominated regime. The table illustrates the dependence of the gravitational coupling $\alpha_{\rm g}$, the gravitational Bohr radius $a_{\rm g}$, and the characteristic radius $R_{\rm c} = \langle r \rangle_{100} = 3a_{\rm g} / 2$ on the axion mass. The hydrogenic approximation requires $\alpha_{\rm g}\ll1$~\cite{Detweiler:1980, Arvanitaki:2010sy, Brito:2015oca}; consequently the $m_{\rm a} = 10^{-10}\,\mathrm{eV}$ entry should be regarded as indicative rather than quantitatively precise.}
\label{tab:solar_BH_cloud}
\end{table}

The condition $\alpha_{\rm g}\ll1$ implies, for a solar-mass black hole, $m_{\rm a} \ll 1.34\times10^{-10}\ {\rm eV}$. Consequently, axion masses in the approximate range $10^{-14}\lesssim m_{\rm a} \lesssim10^{-11}\,\mathrm{eV}$ produce black hole-bound ground states whose characteristic sizes extend from astronomical-unit scales to several hundred kilometres. This hydrogenic solution provides the unperturbed state about which the halo-induced perturbation developed in the main text is constructed.

The corresponding mass-density profile follows directly from the ground-state wave-function,
\begin{equation}
	\rho(r)
		=
			m_{\rm a}|\psi_{100}(r)|^2
		=
			\rho_{\rm c}\,e^{-2r/a_{\rm g}}\,,
	\label{eq:hydrogenic_density_profile}
\end{equation}
where the total condensate mass is
\begin{equation}
	M_{\rm c}
		=
			m_{\rm a}N
		=
			4\pi\int_0^\infty r^2\rho(r)\,{\rm d}r\,.
	\label{eq:cloud_mass_definition}
\end{equation}
Using Eq.~(\ref{eq:hydrogenic_density_profile}) immediately yields
\begin{equation}
	\rho_{\rm c}
		=
			\frac{M_{\rm c}}{\pi a_{\rm g}^3}\,,
	\label{eq:rhoc_ag_final}
\end{equation}
so that the density profile may be written as
\begin{equation}
	\rho(r)
		=
			\frac{M_{\rm c}}{\pi a_{\rm g}^3}e^{-2r/a_{\rm g}}\,.
	\label{eq:hydrogenic_density_profile_final}
\end{equation}

Equation~(\ref{eq:rhoc_ag_final}) highlights an important distinction. While the black hole mass $M_{\rm BH}$ and the scalar mass $m_{\rm a}$ determine the characteristic size of the bound state through the gravitational Bohr radius $a_{\rm g}$, they do not determine the overall normalization of the density profile. The latter is fixed independently by the total mass $M_{\rm c}$ accumulated in the scalar cloud. Consequently, $a_{\rm g}$ sets the spatial extent of the bound state, whereas $M_{\rm c}$ determines its central density.

The corresponding mass-density profile follows directly from the ground-state wavefunction,
\begin{equation}
	R_{\rm c}
		\equiv
			\langle r \rangle_{10}
		=
			\frac{3}{2}a_{\rm g}\,,
	\label{eq:Rc_rhoc_definition}
\end{equation}
so that
\begin{equation}
	a_{\rm g}
		=
			\frac{2}{3}R_{\rm c}\,.
\end{equation}
The relation
\begin{equation}
	\rho_{\rm c}
		=
			\frac{M_{\rm c}}{\pi a_{\rm g}^3}
\end{equation}
can therefore be written as
\begin{equation}
	\rho_{\rm c}
		=
			\frac{27}{8\pi}\frac{M_{\rm c}}{R_{\rm c}^3}
		\simeq
			1.074\,\frac{M_{\rm c}}{R_{\rm c}^3}\,.
	\label{eq:rhoc_Rc_final}
\end{equation}
Equation~(\ref{eq:rhoc_Rc_final}) provides the relation required to assign a central density to a hydrogenic cloud of total mass $M_{\rm c}$ and characteristic radius $R_{\rm c}$. The quantity $\rho_{\rm c}$ denotes the density at the origin and should not be interpreted as the average density inside $R_{\rm c}$.

Substituting the gravitational Bohr radius,
\begin{equation}
	a_{\rm g}
		=
			\frac{\hbar^2}{G M_{\rm BH}m_{\rm a}^2}\,,
\end{equation}
the central density can be expressed directly in terms of the black hole mass, scalar mass, and total cloud mass:
\begin{equation}
	\rho_{\rm c}
		=
			\frac{G^3M_{\rm BH}^3m_{\rm a}^6M_{\rm c}}{\pi\hbar^6}\,.
	\label{eq:rhoc_fundamental}
\end{equation}
For a solar-mass black hole, the gravitational Bohr radius is
\begin{equation}
	a_{\rm g}
		\simeq
			2.64\times10^4\,{\rm km}\left(\frac{M_\odot}{M_{\rm BH}}\right)\!\left(\frac{10^{-12}\ {\rm eV}}{m_{\rm a}}\right)^{\!2}\,,
\end{equation}
corresponding to the characteristic radius
\begin{equation}
	R_{\rm c}
		\simeq
			3.96\times10^4~{\rm km}\left(\frac{M_\odot}{M_{\rm BH}}\right)\!\left(\frac{10^{-12}\ {\rm eV}}{m_{\rm a}}\right)^{\!2}\,.
	\label{eq:Rc_solar_rhoc}
\end{equation}
The associated central density is
\begin{equation}
	\rho_{\rm c}
		\simeq
			5.10\times10^{26}\,\frac{M_\odot}{{\rm pc}^3}
            \left(
                \frac{M_{\rm c}}{M_\odot}
            \right)\!
            \left(
                \frac{M_{\rm BH}}{M_\odot}
            \right)^{\!3}
            \left(
                \frac{m_{\rm a}}{10^{-12}\ {\rm eV}}
            \right)^{\!6}
            \,.
	\label{eq:rhoc_solar_astro}
\end{equation}
The enclosed halo mass within the characteristic radius \(R_{\rm c}\) is
\begin{equation}
    M( < R_{\rm c})
        =
            4\pi\int_0^{R_{\rm c}}\rho(r)\,r^2\,{\rm d}r
        =
            4\pi\rho_{\rm c}
            \int_{0}^{R_{\rm c}}
            \frac{\sin(kr)}{kr}\,r^2\,P{\rm d}r
        =
            \frac{4}{\pi}\.\.\rho_{\rm c} R_{\rm c}^3
            \, .
\end{equation}
We restrict the perturbative analysis to configurations satisfying
\begin{equation}\label{eq:min_halo_req}
    \frac{M_{\rm halo}(<r_{\rm char})}{M_{\rm BH}}
        \lesssim
                10^{-2}
                \, ,
\end{equation}
ensuring that the halo contribution remains subdominant over the characteristic support of the bound state. Under this criterion, higher-order corrections are expected to remain at the few-percent level.

\begin{figure}[!h]
\begin{minipage}{0.43\textwidth}
    \includegraphics[width=1.0\linewidth]{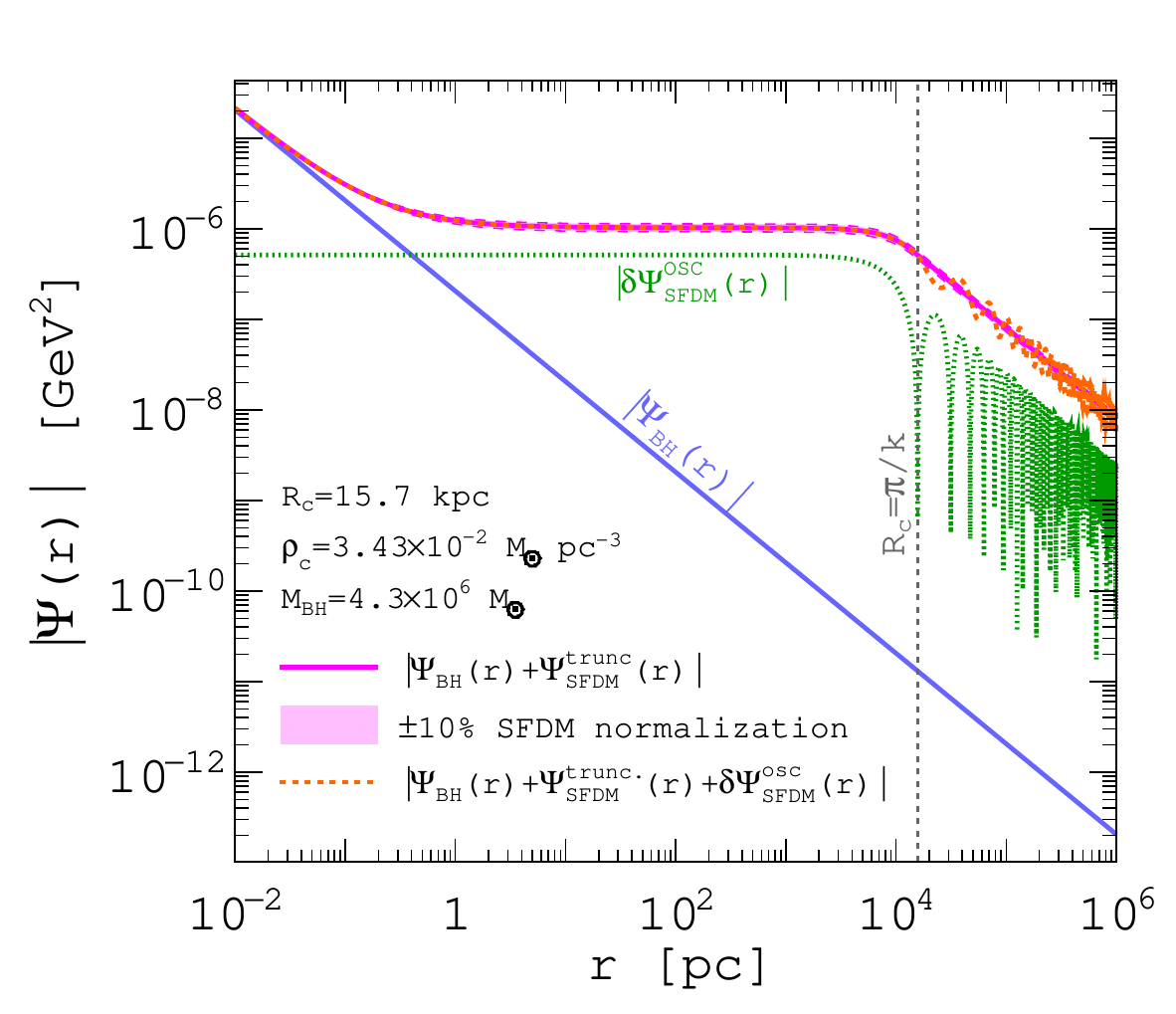}
    \end{minipage}
    \hspace{0.5 cm}
    \begin{minipage}{0.43\textwidth}
    \includegraphics[width=1.0\linewidth]{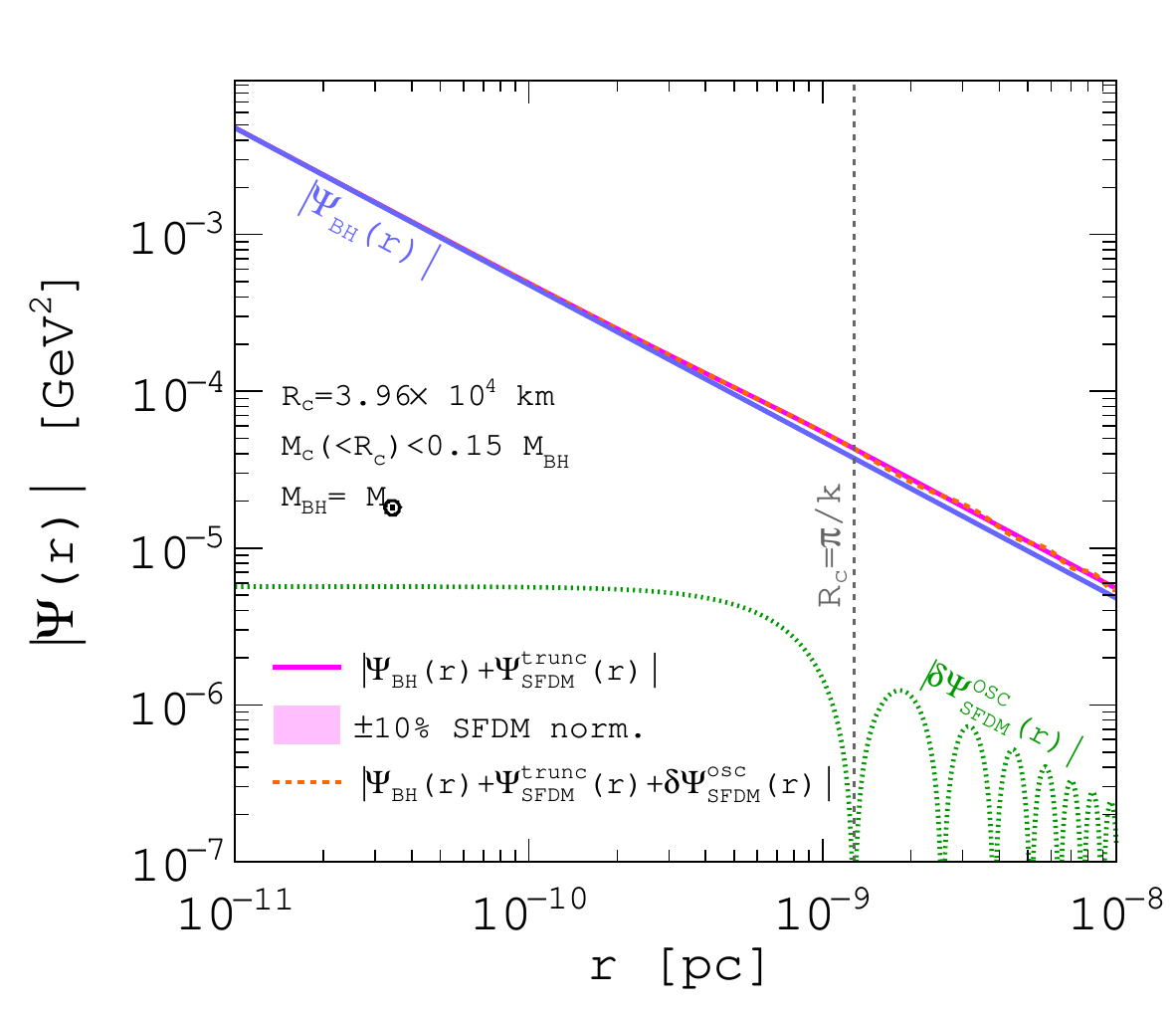}
    \end{minipage}
\caption{
Comparison of the gravitational potential generated by a Bose--Einstein-condensed scalar-field dark matter halo surrounding (left) a supermassive black hole representative of the Galactic centre and (right) a stellar-mass black hole. In both panels, the blue curve shows the Schwarzschild potential, while the solid magenta curve corresponds to the total gravitational potential obtained by adding the physically truncated halo contribution. The shaded band indicates the effect of a $\pm 10\%$ uncertainty in the halo normalization. The vertical dashed line marks the characteristic halo radius, $R_{\rm c} = \pi/k$, where the interior BEC solution is matched to the exterior $1/r$ potential. For comparison, the green dotted curve shows the oscillatory component of the halo potential, $\big| \delta\Psi_{\rm SFDM}^{\rm osc}(r) \big|$, and the orange dashed curve illustrates the corresponding oscillatory continuation of the total potential beyond $R_{\rm c}$. The stellar-mass benchmark is chosen such that the enclosed halo mass satisfies $M_{\rm halo}( < R_{\rm c} ) < 10^{-2}\,M_{\rm BH}$, ensuring that the perturbative treatment adopted throughout this work remains valid.
}
\label{fig:framework}
\end{figure}

\section{Additional information regarding the simulation tools}

\subsection{General Information}

The code used for the geodesic simulations is an implementation of the numerical algorithm presented in~\cite{Bacchini:2018} and has already been used to simulate particles around rotating neutron stars~\cite{Cristoph}. Trajectories of both massless and massive particles can be simulated, regardless of the actual value of their mass, since the equivalence principle states that each particle, regardless of its mass, follows a geodesic~\cite{Tong:2019}.

\begin{figure}[!h]
    \centering
    \includegraphics[width=0.35\linewidth]{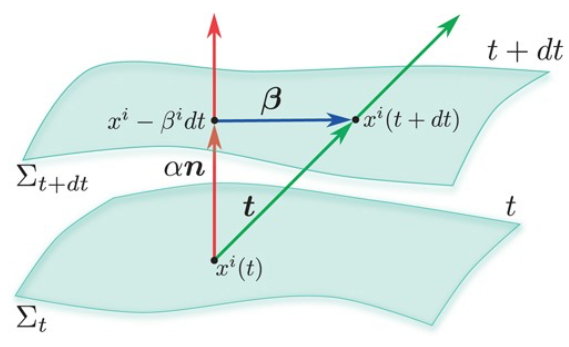}
    \caption{3+1 foliation of spacetime into spacelike hypersurfaces, where \textbf{\textit{t}} is a 4-vector representing the evolution of the time coordinate $t$. \textbf{\textit{t}} can be split into a timelike component, $\alpha\.\.$\textbf{\textit{n}}, and a spacelike one, $\boldsymbol{\beta}$: $\alpha$ measures the proper time between adjacent hypersurfaces, \textbf{\textit{n}} is a timelike unit normal to the hypersurface and $\boldsymbol{\beta}$ measures the change of coordinates from one hypersurface to the other~\cite{Rezzolla:2013}.}
    \label{fig:ADM-formalism-foliation}
\end{figure}

The geodesic equation describes the motion of particles in curved spacetime
\begin{equation} \label{eq:geodesics}
    \frac{{\rm d}^2 x^\mu}{{{\rm d} \tau}^2} + \Gamma_{\nu \lambda}^\mu \frac{{\rm d} x^\nu}{{\rm d} \tau} \frac{{\rm d} x^\lambda}{{\rm d} \tau}
        =
            0
            \, ,
    \qquad \qquad
    \Gamma_{\nu \lambda}^\mu
        =
            \frac{1}{2} g^{\mu \sigma}
            \big( 
                \partial_\lambda g_{\sigma \nu} + \partial_\nu g_{\sigma \lambda} - \partial_{\sigma} g_{\nu \lambda}
            \big)
            \, ,
\end{equation}
where $\mu = 0, \, 1, \, 2, \, 3$, $\Gamma_{\nu \lambda}^\mu$ are the Christoffel symbols and $\tau$ is an affine parameter. In numerical integration it is common to use the Arnowitt--Deser--Misner (ADM) formalism, which makes use of a 3+1 spacetime decomposition into spacelike hypersurfaces (Fig.~\ref{fig:ADM-formalism-foliation}). 

Any metric can be expressed according to this 3+1 foliation:
\begin{equation} \label{eq:ADM-formalism-metric}
    g_{\mu \nu} 
        = 
            \left(
                \begin{array}{cc}
                    - \alpha^2 + \beta_{k}\. \beta^{k} & \beta_{i} \\
                    \beta_{j} & \gamma_{ij} \\
                \end{array}
                \right)
    \qquad \qquad 
    g^{\mu \nu}
        = 
            \left(
                \begin{array}{cc}
                    - 1 / \alpha^2  & \beta^{i} / \alpha^2 \\
                    \beta^{j} / \alpha^2 & \gamma^{ij} - \beta^{i} \beta^{j} / \alpha^2 \\
                \end{array}
            \right)
                ,
\end{equation}
where $\alpha$ is the so-called ``lapse function'', $\beta_{i}$ the ``shift vector'' and $\gamma_{ij}$ is the spatial part of $g_{\mu \nu}$ and their expressions have been derived from the scalar field dark matter (SFDM)~\cite{Suarez2014} halo metric as can be seen from the expression of the line element in ADM formalism (with $(-,+,+,+)$ signature)
\begin{equation} \label{eq:line-element-ADM-formalism}
    {\rm d}s^2
        =
            - \alpha^2 {\rm d}t^2 + \gamma_{ij} \big({\rm d}x^i + \beta^{i} {\rm d}t)({\rm d}x^j + \beta^{j} {\rm d}t\big)
            \, .
\end{equation}

By defining $u^\mu \equiv {\rm d} x^\mu / {\rm d}\tau$ and $u_i \equiv g_{i \mu} u^\mu$, Eq.~\eqref{eq:geodesics} can be rewritten in terms of Eqs.~\eqref{eq:evolution-eq-first} and ~\eqref{eq:evolution-eq-second}
\begin{equation} \label{eq:evolution-eq-first}
    \frac{{\rm d}x^i}{{\rm d}t}
        =
            \gamma^{ij} \frac{u_j}{u^{0}} - \beta^{i}
            \, ,
\end{equation}
\begin{equation} \label{eq:evolution-eq-second}
    \frac{{\rm d} u_i}{{\rm d}t}
        =
            - \alpha u^{0} \partial_i \alpha + u_k \partial_i \beta^{k} - \frac{u_j u_k}{2 u^{0}} \partial_i \gamma^{jk}
            \, ,
\end{equation}
where the definition of $u^{0}$ [see Eq.~\eqref{eq:evolution-eq-third}] helps reduce the complexity of the numerical scheme because the integration of the temporal components $x^{0}$ and $u^{0}$ is no longer needed.

\begin{equation} \label{eq:evolution-eq-third}
    u^{0}
        =
            \frac{1}{\alpha} \, \sqrt{\gamma^{jk} u_j u_k + \epsilon}
            \; ,
    \qquad\qquad
    \epsilon =
            \bigg\{
            \begin{array}{rl}
                1 & \qquad \mathrm{(massive \> particles)}\\
                0 & \qquad \mathrm{(massless \> particles)} \\
            \end{array}
        \, .
\end{equation}

The Hamiltonian for stationary metrics is then defined as~\cite{Bacchini:2018} 
\begin{equation} \label{eq:hamiltonian}
    H\big( x^i, u^i \big)
        =
            \alpha \sqrt{\gamma^{ij} u_j u_k + \epsilon} - \beta^{j} u_j
            \, ,
\end{equation}
and one can show that the Killing equation (the Killing vector is $K = \partial_t$) implies that $K^\mu u_\mu = u_0 = - \alpha^2 u^{0} + \beta^{j} u_j$ is conserved. Using Eq.~\eqref{eq:evolution-eq-third}, one can retrieve that $|H| = |u_0|$ which implies that a numerical scheme that conserves the Hamiltonian also conserves the energy, this is why the algorithm this code is based on is said to be an ``energy-conserving scheme''. For the discretisation of the equations from which the equations of motion are derived, namely ${\rm d}x^i / {\rm d}t = \partial H(x^i, u^i) / \partial u^i$ and ${\rm d}u^i / {\rm d}t = - \partial H(x^i, u^i) / \partial x^i$, refer to the Appendix of~\cite{Bacchini:2018}.

\subsection{Simulation Script}

The simulation script is written in \texttt{Python} and, in order to run, it needs the name of the metric (the metric parameters are hard-coded inside the main script), the celestial body spin and dimensionless reduced quadrupole (for rotating neutron stars), a boolean that has the same role of $\epsilon$ in Eq.~\eqref{eq:evolution-eq-third}, the initial conditions on position and speed, the simulation time step and number of time steps (these two parameters define the maximum time of the simulation), and a string that represents the path of the directory where the output will be saved. 

Across the entire simulation geometrised ($G=c=1$) black hole ($M_{\rm BH} = 1$) units are used. Quantities in real units can be retrieved through Eq.~\eqref{eq:unit-conversion-formulas}
\begin{equation} \label{eq:unit-conversion-formulas}
    R
        =
            R^{\mathrm{sim}}\,\frac{ r_{\rm g} }{ r_{\rm g}^{\mathrm{sim}} }
            \, ,
    \qquad \qquad
    t
        =
            t^{\mathrm{sim}}\,\frac{ t_{\rm g} }{ t_{\rm g}^{\mathrm{sim}} }
            \, ,
\end{equation}
where $r_{\rm g}$ is the gravitational radius and $t_{\rm g}$ is the gravitational time (see below for definitions), while $r_{\rm g}^{\mathrm{sim}}$ and $t_{\rm g}^{\mathrm{sim}}$ are their equivalent in simulation units, both normalised to 1 by convention. 
\begin{equation} \label{eq:gravitational-radius-time}
    r_{\rm g}
        =
            \frac{ G\,M_{\rm BH} }{ c^2 }\, ,
    \qquad\qquad
    t_{\rm g}
        =
            \frac{ G\,M_{\rm BH} }{ c^3 }
            \, .
\end{equation}

In order to obtain the initial conditions for the trajectories, given that Ref.~\cite{Levin:2008} provides the initial energy and angular momentum only, in simulations where Schwarzschild black holes are present, the initial position was initialized with one of the roots of Eq.~\eqref{eq:polynomial-Schwarzschild}
\begin{equation}
    \label{eq:polynomial-Schwarzschild}
    E^2\,r^4 - \big( r^2 - r_{\rm s}\,r \big)
    \big( r^2 - L^2 \big)
        =
            0
            \, .
\end{equation}
The polar angle $\theta$ has been set to $\pi / 2$ (so that the trajectory would be on the equatorial plane), $\phi$ has been set equal to 0, as $u_{r}$ and $u_{\theta}$ also have, while $u_{\phi}$ has been set equal to the initial angular momentum $L$.

To demonstrate the scalability of the framework across a wide range
of masses and environments, we consider the five benchmark
configurations summarized in Table~II. The first three correspond to
compact SFDM clouds surrounding black holes of masses
$1\,M_\odot$, $10\,M_\odot$, and $4.3\times10^{6}\,M_\odot$,
respectively. The remaining two configurations describe supermassive
black holes embedded in extended galactic SFDM halos, corresponding
to the Galactic Centre and ESO~120-0211. The halo parameters are
chosen such that Eq.~(S100) is satisfied, ensuring that the spacetime
remains perturbatively close to the vacuum solution. For the
compact-cloud benchmarks, we adopt the conservative choice
$M_{\rm c}=M_{\rm halo}(<r_{\rm char})\simeq10^{-5}M_{\rm BH}$,
well below the perturbative upper bound implied by Eq. \eqref{eq:min_halo_req}.
This choice ensures that the orbital modifications arise from the
weak gravitational influence of the surrounding SFDM cloud rather
than from a significant redistribution of the total gravitational
mass.

\begin{table*}[!h]
\centering
\renewcommand{\arraystretch}{1.35}
\setlength{\tabcolsep}{9pt}
\begin{tabular}{llccccc}
\cline{3-7}
&
&
\multicolumn{2}{c}{Stellar-mass black holes}
&
\multicolumn{3}{c}{Supermassive black holes}
\\
\cline{3-7}
&
&
$1\,M_{\odot}$
&
$10\,M_{\odot}$
&
$4.3\times10^{6}\,M_{\odot}$
&
$4.3\times10^{6}\,M_{\odot}$
&
$5.62\times10^{6}\,M_{\odot}$
\\
&
&
Compact Cloud
&
Compact Cloud
&
Compact Cloud
&
GC Halo
&
ESO Halo
\\
\hline
\hline

\multirow{5}{*}{\rotatebox[origin=c]{90}{Cloud}}
&
$\rho_{\rm c}\,[M_{\odot}\,{\rm pc}^{-3}]$
&
$5.10\times10^{21}$
&
$5.10\times10^{25}$
&
$1.6\times10^{18}$
&
$3.43\times10^{-2}$
&
$1.37\times 10^{-2}$
\\
&
$R_{\rm c}$ [km]
&
$3.96\times10^{4}$
&
$3.96\times10^{3}$
&
$8.64\times 10^{7}$
&
$4.84\times 10^{17}$
&
$9.01\times 10^{16}$
\\
&
$M(<R_{\rm c})\,[M_{\odot}]$
&
$1.37\times10^{-5}$
&
$1.37\times10^{-4}$
&
$44.8$
&
$1.69\times10^{11}$
&
$4.34\times 10^{8}$
\\
&
$m_a\,[{\rm eV}]$
&
$4.28\times 10^{-11}$
&
$4.28\times 10^{-12}$
&
$10^{-17}$
&
$10^{-17}$
&
$7.6\times 10^{-18}$
\\
&
$\langle r\rangle_{10}^{(0)}$ [km]
&
$21.63$
&
$216.3$
&
$9.21\times10^{7}$
&
$9.21\times10^{7}$
&
$1.22\times 10^{8}$
\\
\hline
\multirow{2}{*}{\rotatebox[origin=c]{90}{Schw.}}
&
$r_{\rm char}$ [km]
&
$19.925$
&
$199.249$
&
$8.756\times 10^{7}$
&
$8.630\times10^{7}$ 
&
$1.119\times10^{8}$
\\
&
$\delta_r$ [\%]    
& $-7.88$ 
& $-7.88$ 
& $-4.93$
& $-6.30$
& $-8.28$ \\
\hline

\multirow{2}{*}{\rotatebox[origin=c]{90}{Kerr}}
&
$r_{\rm char}$ [km]
&
$16.570$
&
$165.468$
&
$7.114\times 10^{7}$
&
$7.114\times 10^{7}$
&
$9.313\times10^{7}$
\\
&
$\delta_r$ [\%]     
& $-23.39$ 
& $-23.50$ 
& $-22.76$ 
& $-22.76$
& $-23.66$
\\
\hline
\hline

\end{tabular}
\caption{Benchmark parameters for the numerical simulations. The first three columns correspond to compact SFDM clouds with $M_{\rm c}=10^{-5}M_{\rm BH}$, while the last two describe the Galactic-Center (GC) and ESO 120-0211 galactic halos. The cloud block lists $\rho_{\rm c}$, $R_{\rm c}$, $M(<R_{\rm c})$, $m_a$, and the hydrogenic ground-state radius $\langle r\rangle_{10}^{(0)}=3a_{\rm g}/2$, evaluated for $\alpha_{\rm g}=0.32$. For Schwarzschild (Kerr), $E=0.988\,(0.977)$ and $L=3.9\,(3.63)$, with $a_\ast=0.2$ for Kerr. Here, $\delta_r\equiv[r_{\rm char}-\langle r\rangle_{10}^{(0)}]/\langle r\rangle_{10}^{(0)}\times100\%$. Higher-precision inputs are given in~\cite{Mungai2026}.}
\label{tab:simulation_parameters}
\end{table*}

\begin{figure*}[p]
\centering

\begin{minipage}{0.45\textwidth}
    \centering
    \includegraphics[width=\linewidth]{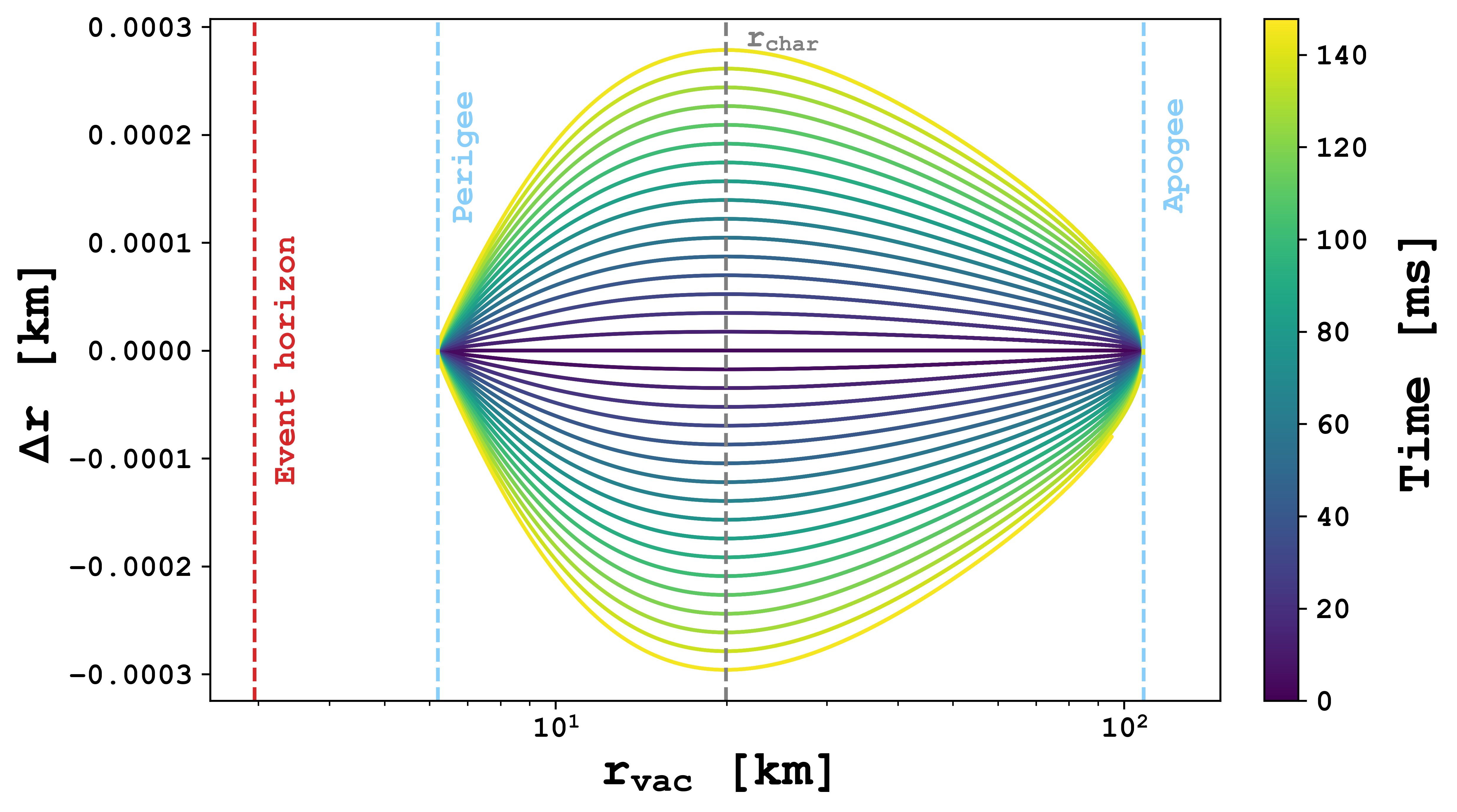}
    \par\smallskip\centering\small $M_{\rm BH}=1\,M_{\odot}$ (Schwarzschild \& Compact Cloud).\par
\end{minipage}
\hfill
\begin{minipage}{0.45\textwidth}
    \centering
    \includegraphics[width=\linewidth]{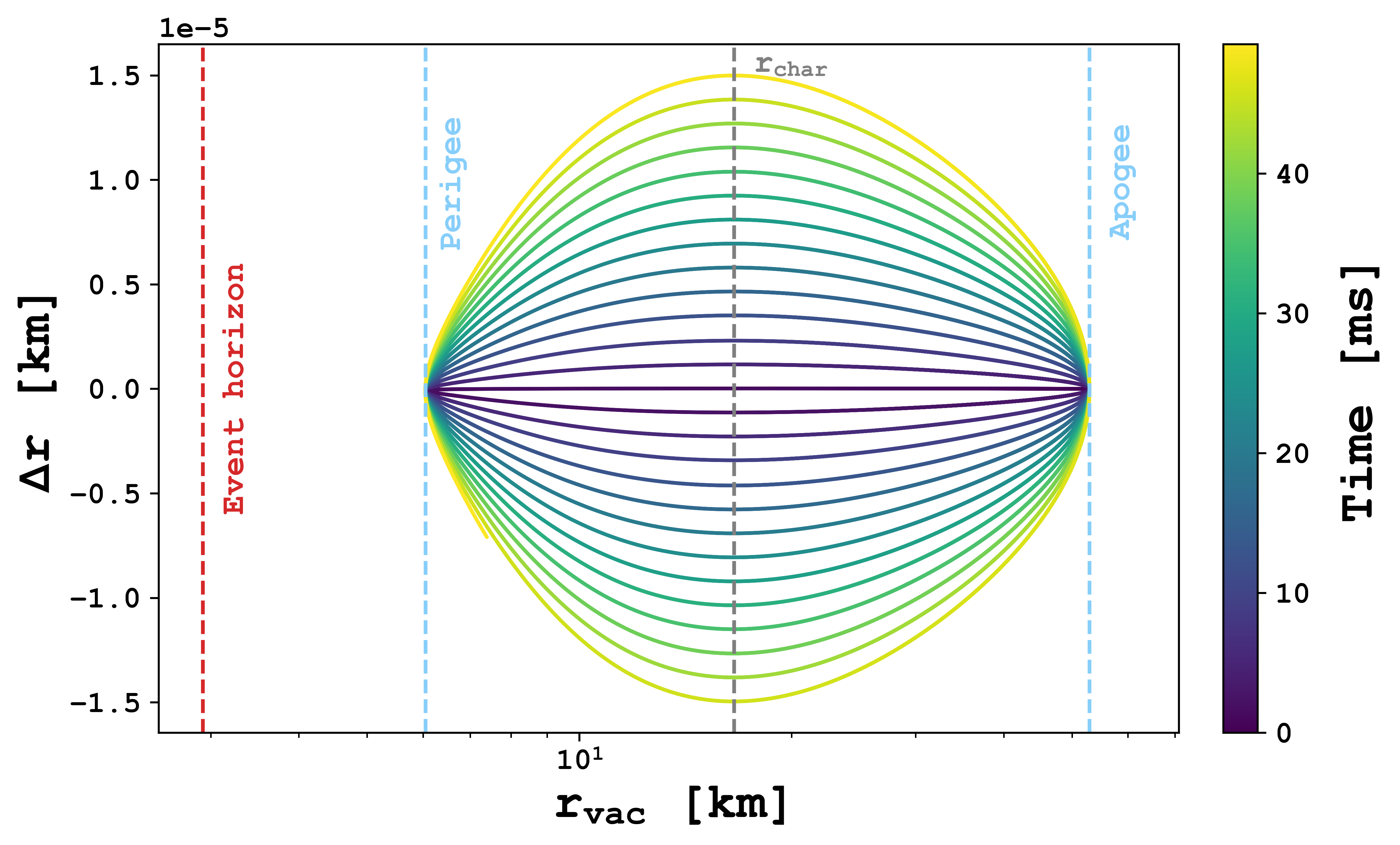}
    \par\smallskip\centering\small $M_{\rm BH}=1\,M_{\odot}$, $a_{\ast}=0.2$ (Kerr \& Compact Cloud).\par
\end{minipage}

\begin{minipage}{0.45\textwidth}
    \centering
    \includegraphics[width=\linewidth]{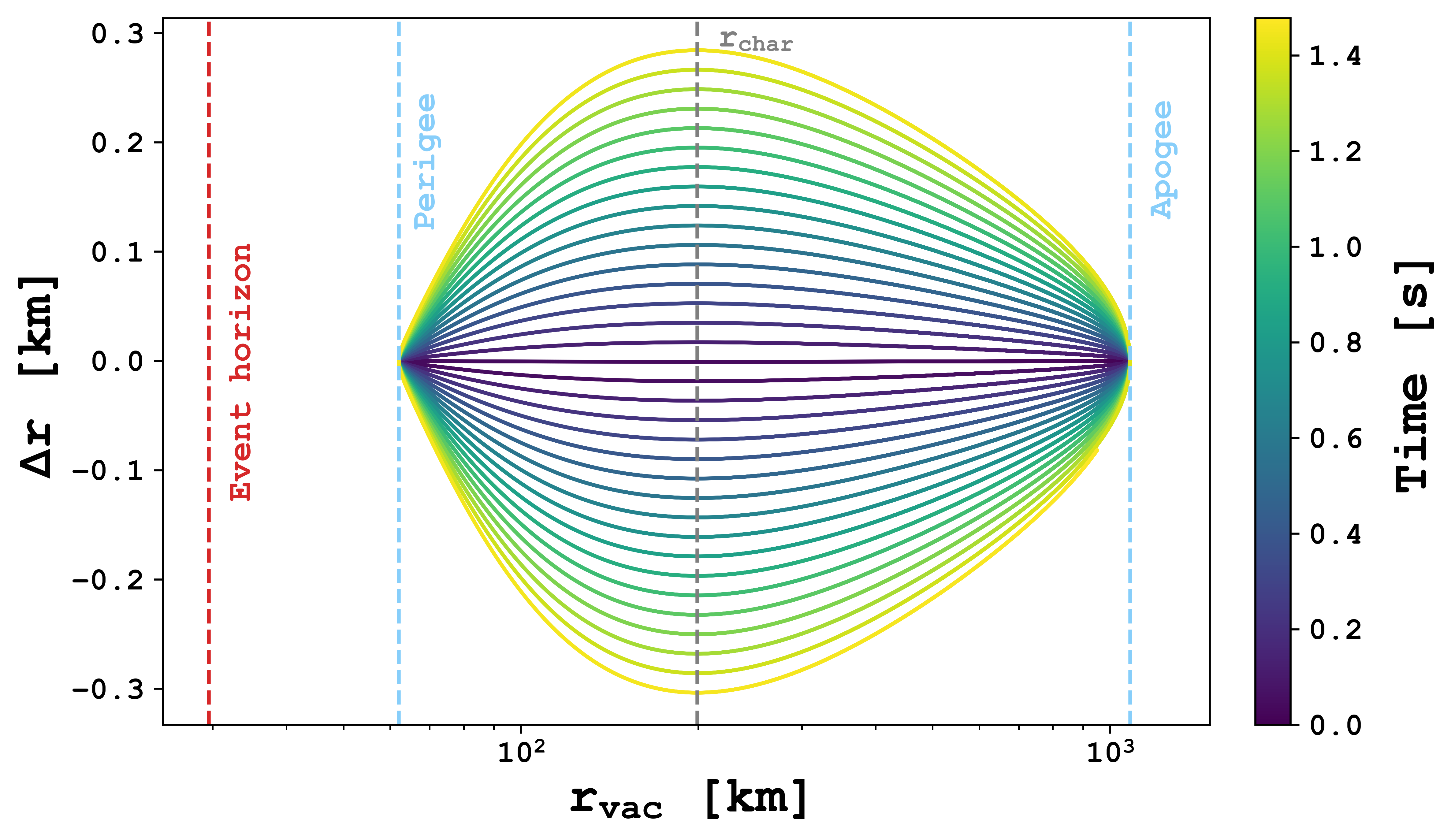}
    \par\smallskip\centering\small $M_{\rm BH}=10\,M_{\odot}$ (Schwarzschild \& Compact Cloud).\par
\end{minipage}
\hfill
\begin{minipage}{0.45\textwidth}
    \centering
    \includegraphics[width=\linewidth]{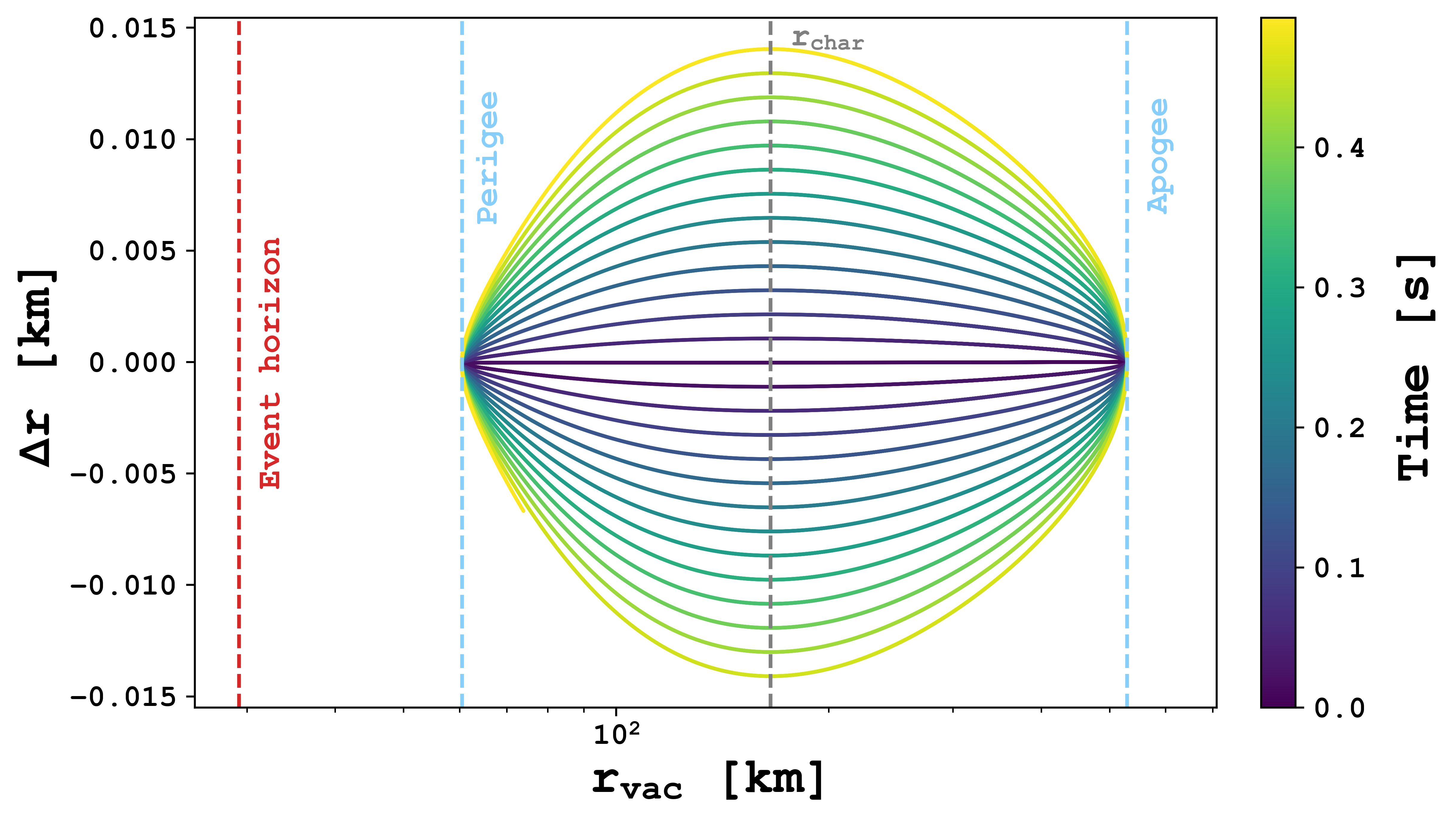}
    \par\smallskip\centering\small $M_{\rm BH}=10\,M_{\odot}$, $a_{\ast}=0.2$ (Kerr \& Compact Cloud).\par
\end{minipage}

\begin{minipage}{0.45\textwidth}
    \centering
    \includegraphics[width=\linewidth]{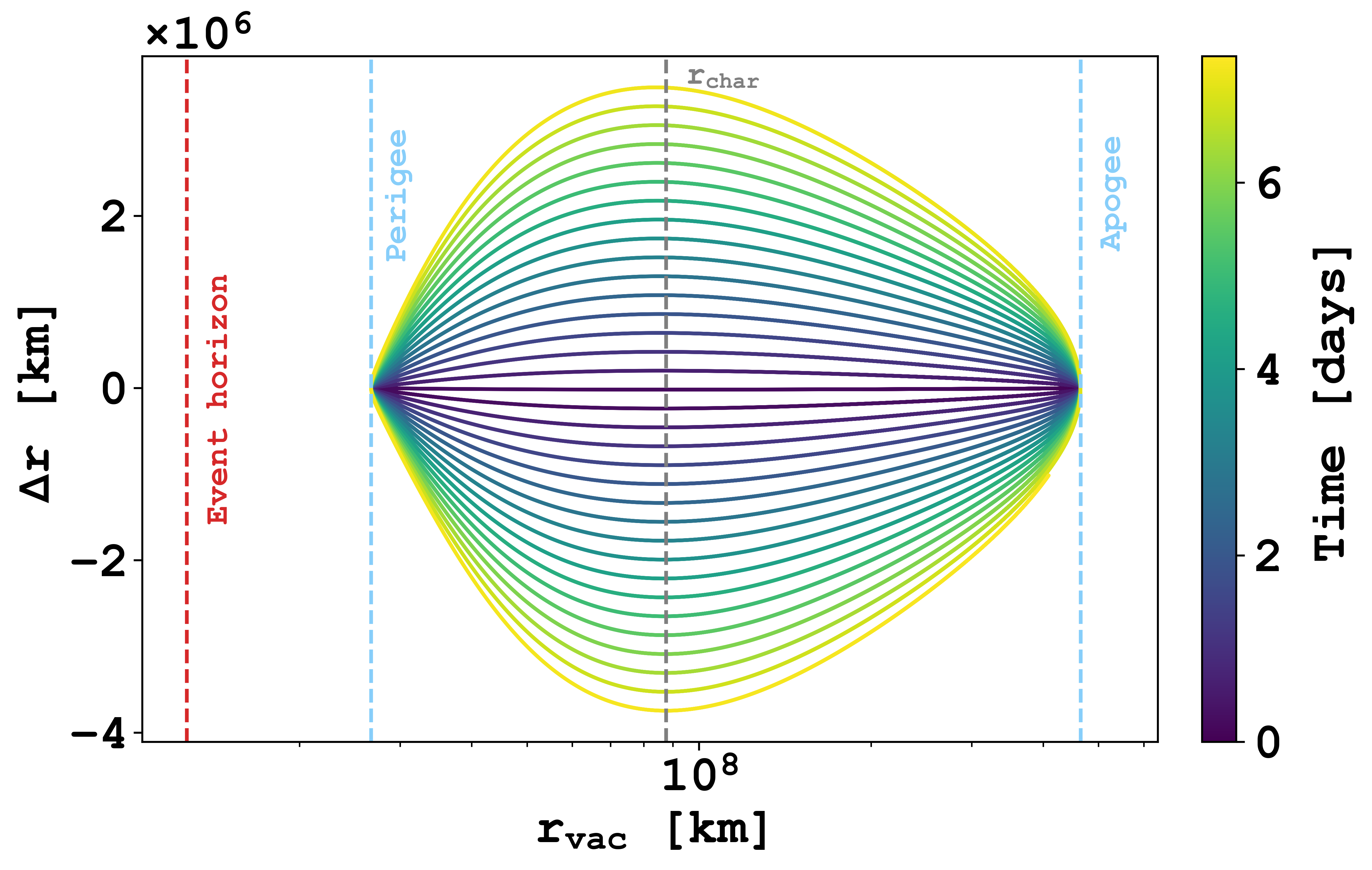}
    \par\smallskip\centering\small Supermassive BH (Schwarzschild \& Compact Cloud).\par
\end{minipage}
\hfill
\begin{minipage}{0.45\textwidth}
    \centering
    \includegraphics[width=\linewidth]{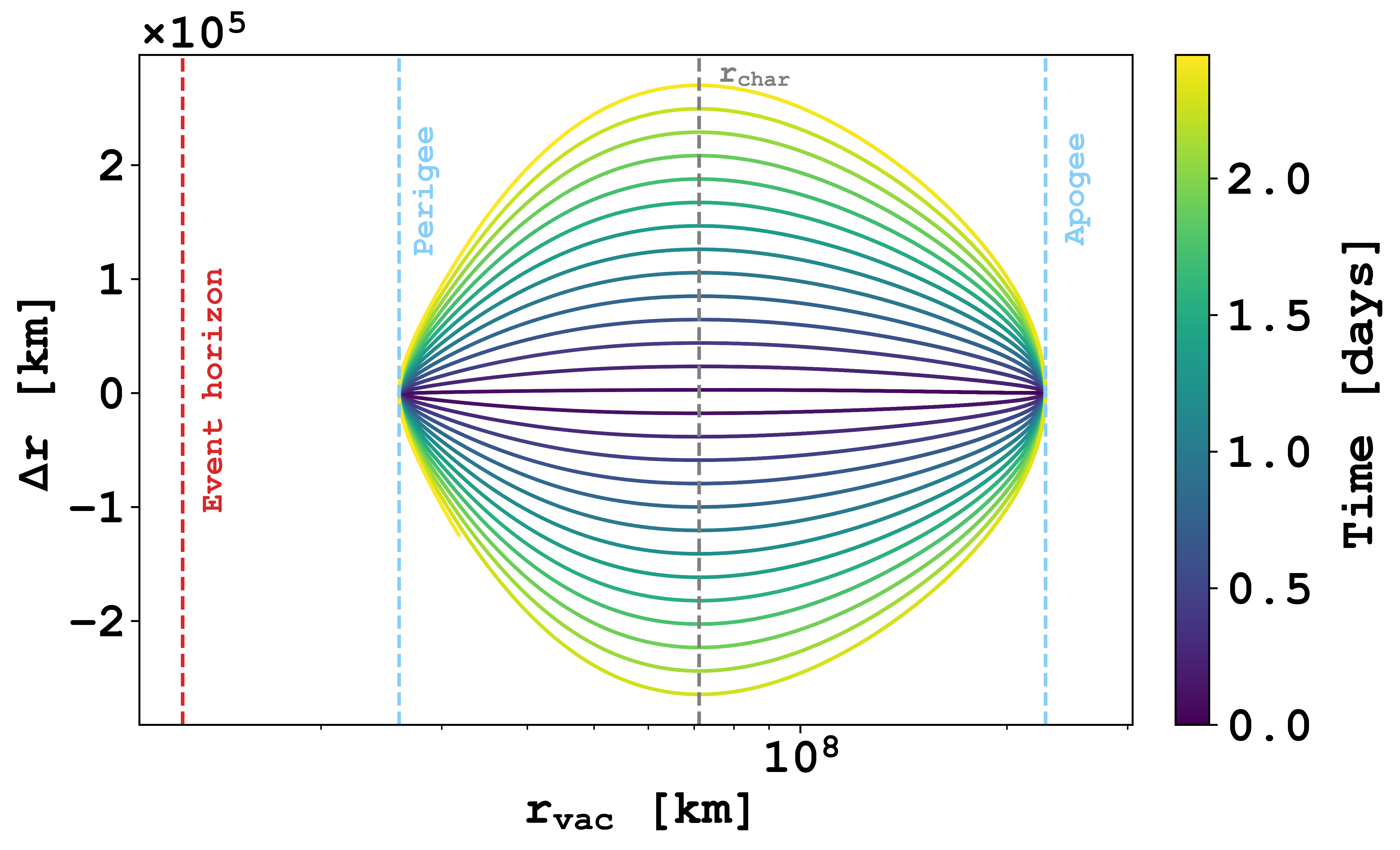}
    \par\smallskip\centering\small Supermassive BH, $a_{\ast}=0.2$ (Kerr \& Compact Cloud).\par
\end{minipage}

\begin{minipage}{0.45\textwidth}
    \centering
    \includegraphics[width=\linewidth]{schwarzschild-4leaf-GC-TIME-day_FIX.png}
    \par\smallskip\centering\small Supermassive BH (Schwarzschild \& Galactic Halo).\par
\end{minipage}
\hfill
\begin{minipage}{0.45\textwidth}
    \centering
    \includegraphics[width=\linewidth]{kerr-4leaf-GC-TIME-day_FIX.png}
    \par\smallskip\centering\small Supermassive BH, $a_{\ast}=0.2$ (Kerr \& Galactic Halo).\par
\end{minipage}

\caption{Radial displacement $\Delta r=r_{\rm SFDM}-r_{\rm vac}$ as a function of the corresponding vacuum orbital radius $r_{\rm vac}$ for representative black-hole masses. The left (right) column shows Schwarzschild (Kerr) spacetimes. The first three rows correspond to black holes of $1\,M_{\odot}$, $10\,M_{\odot}$, and a supermassive black hole of $4.3\times10^{6}\,M_{\odot}$ each surrounded by a compact SFDM cloud with total mass $M_{\rm c}=10^{-5}M_{\rm BH}$. The fourth row instead shows the Galactic-center supermassive black hole embedded in a Galactic-scale SFDM halo. The vertical dashed line marks the characteristic radius $r_{\rm char}$, defined by the maximum of $\Delta r$. Kerr cases assume a moderate spin $a_{\ast}=0.2$.}
\label{fig:simulation_block1}
\end{figure*}

\begin{figure*}[p]
\centering

\begin{minipage}{0.45\textwidth}
    \centering
    \includegraphics[width=\linewidth]{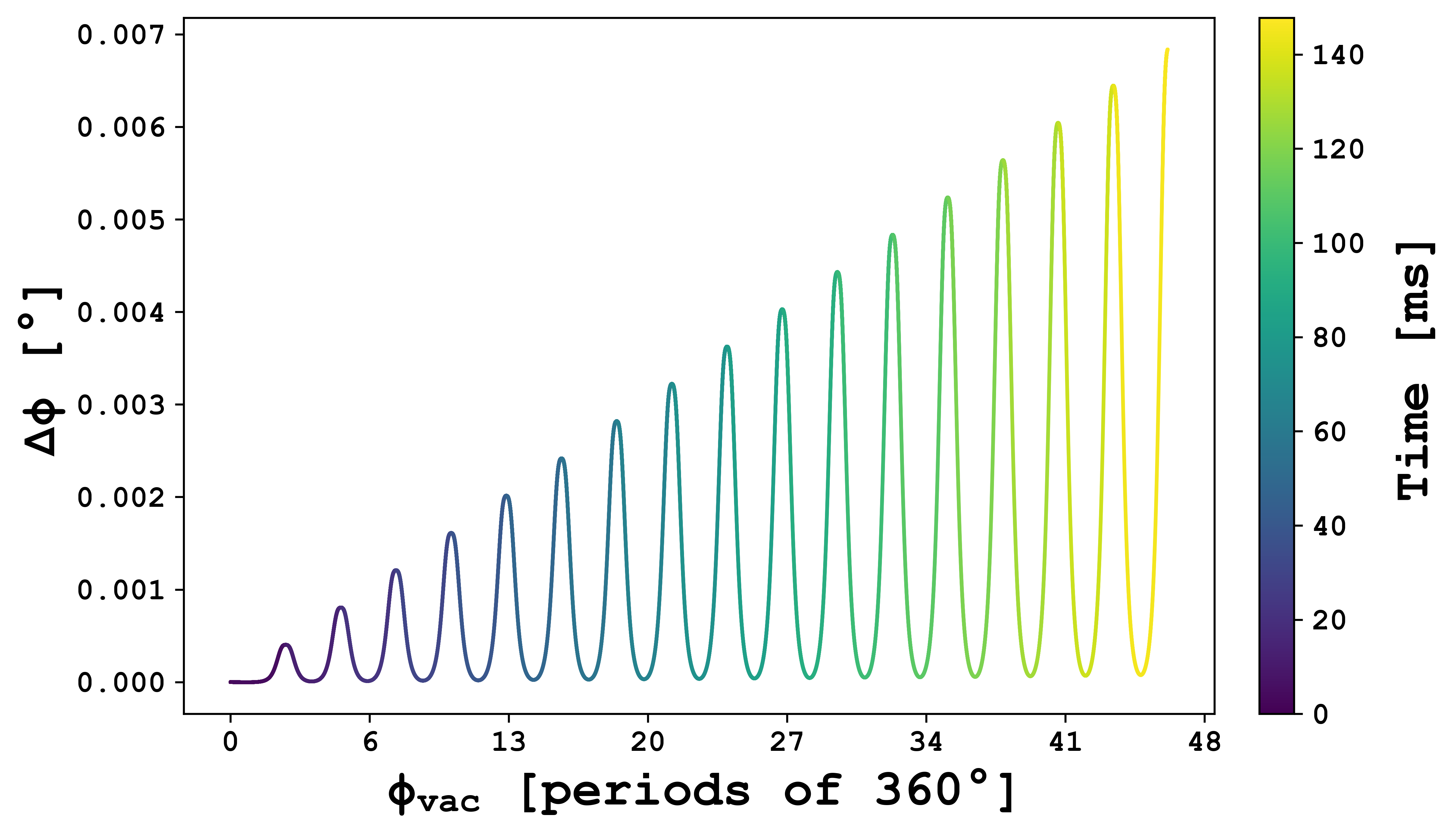}
    \par\smallskip\centering\small $M_{\rm BH}=1\,M_{\odot}$ (Schwarzschild \& Compact Cloud).\par
\end{minipage}
\hfill
\begin{minipage}{0.45\textwidth}
    \centering
    \includegraphics[width=\linewidth]{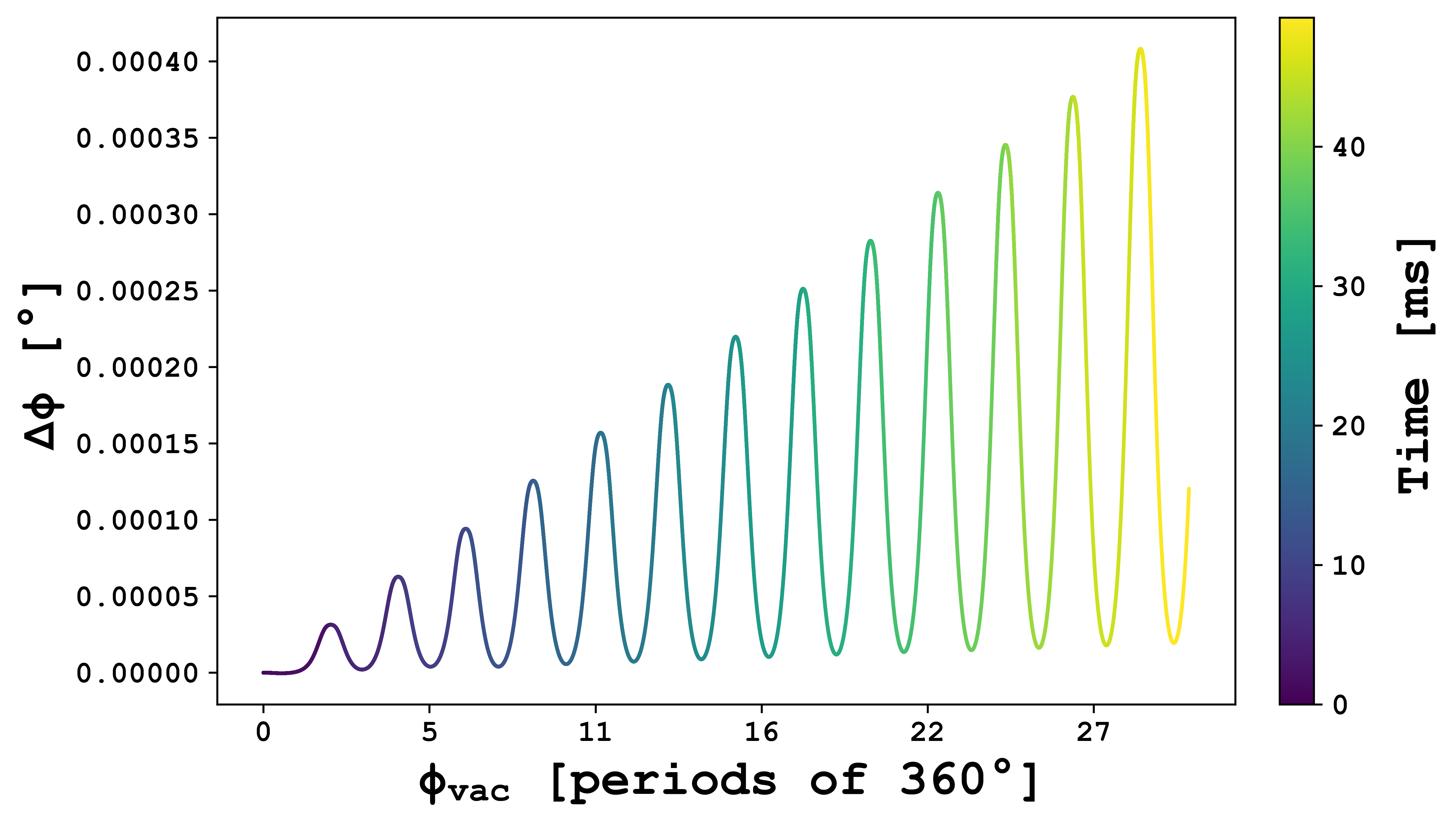}
    \par\smallskip\centering\small $M_{\rm BH}=1\,M_{\odot}$, $a_{\ast}=0.2$ (Kerr \& Compact Cloud).\par
\end{minipage}

\begin{minipage}{0.45\textwidth}
    \centering
    \includegraphics[width=\linewidth]{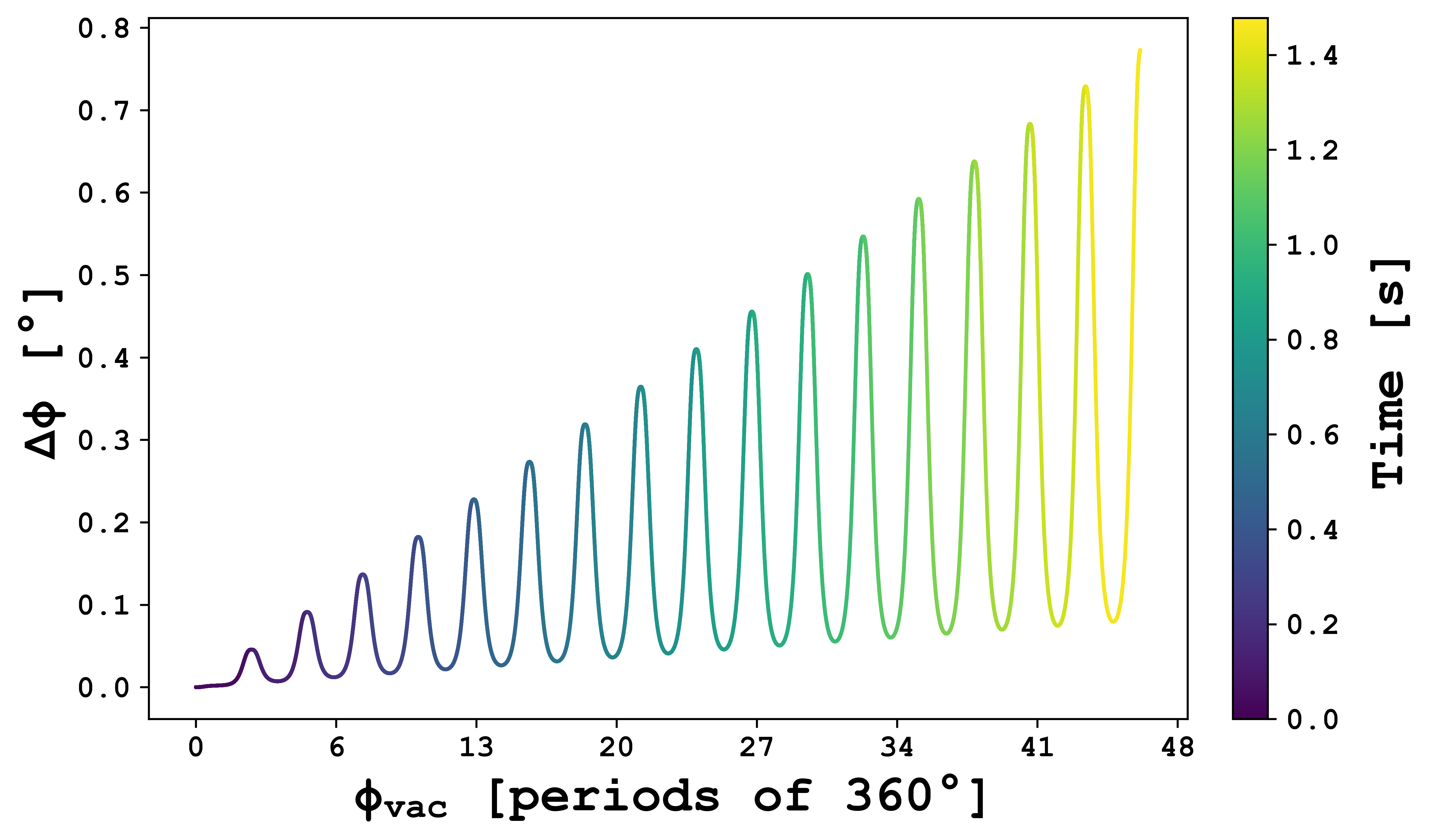}
    \par\smallskip\centering\small $M_{\rm BH}=10\,M_{\odot}$ (Schwarzschild \& Compact Cloud).\par
\end{minipage}
\hfill
\begin{minipage}{0.45\textwidth}
    \centering
    \includegraphics[width=\linewidth]{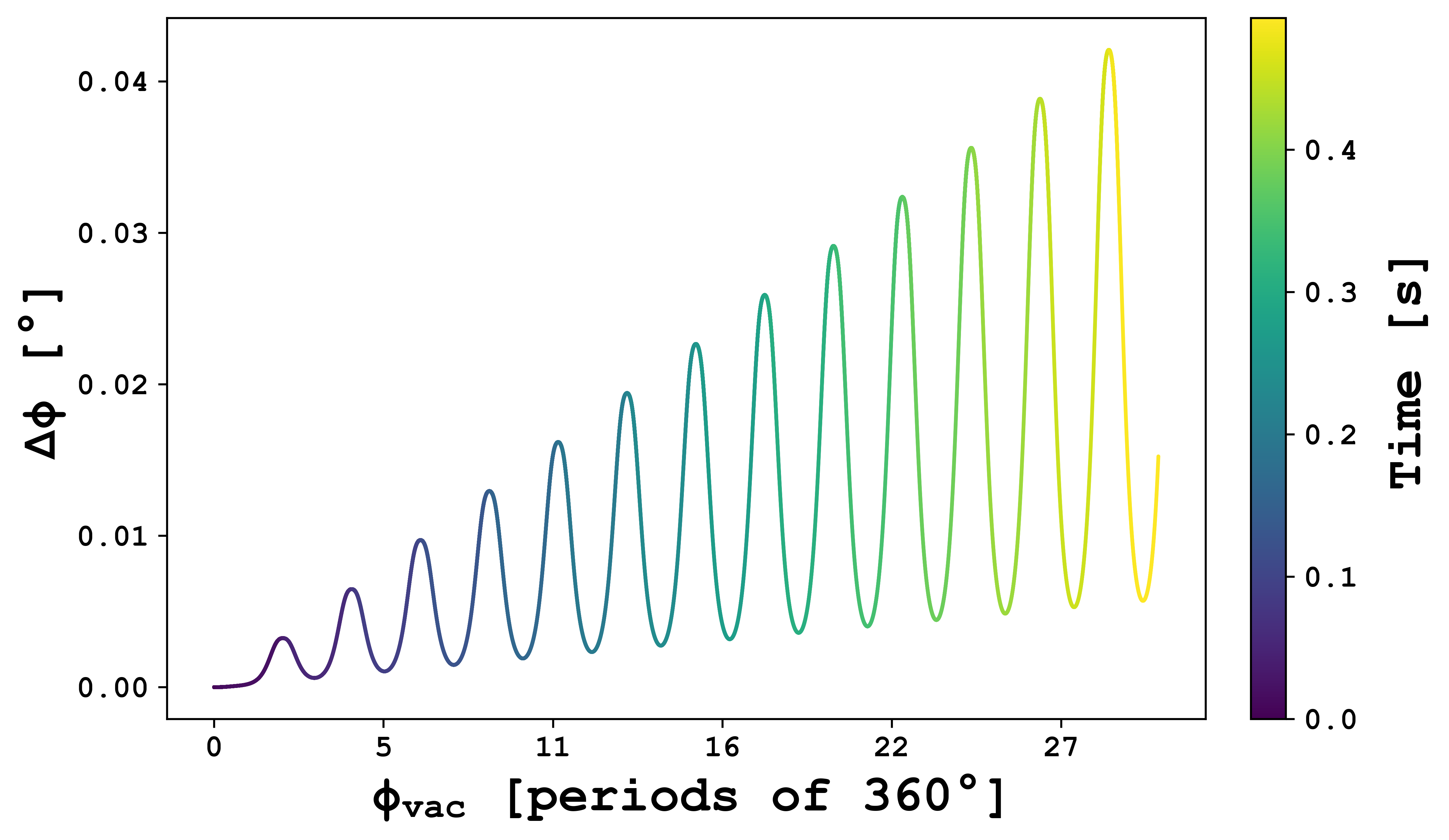}
    \par\smallskip\centering\small $M_{\rm BH}=10\,M_{\odot}$, $a_{\ast}=0.2$ (Kerr \& Compact Cloud).\par
\end{minipage}

\begin{minipage}{0.45\textwidth}
    \centering
    \includegraphics[width=\linewidth]{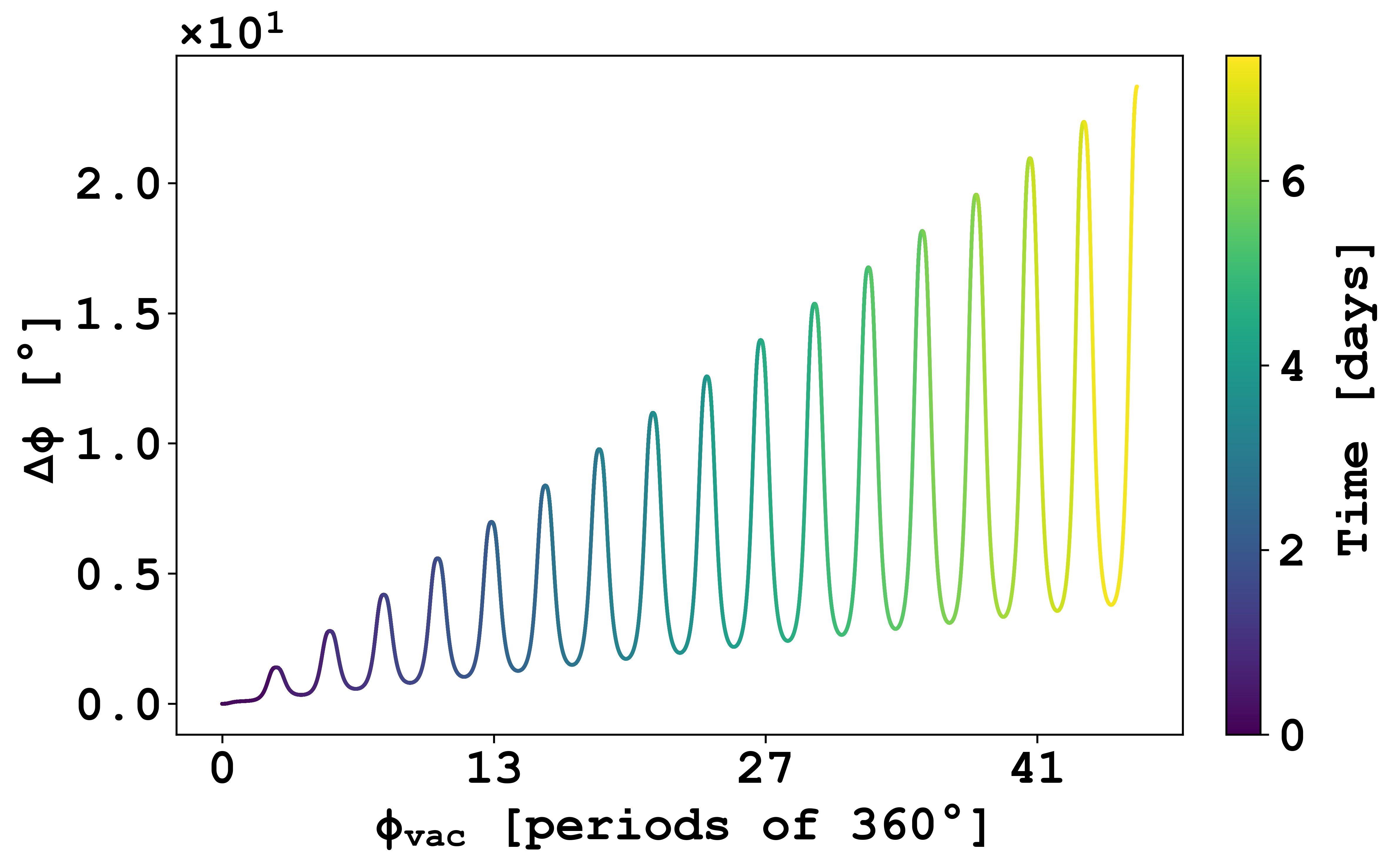}
    \par\smallskip\centering\small Supermassive BH (Schwarzschild \& Compact Cloud).\par
\end{minipage}
\hfill
\begin{minipage}{0.45\textwidth}
    \centering
    \includegraphics[width=\linewidth]{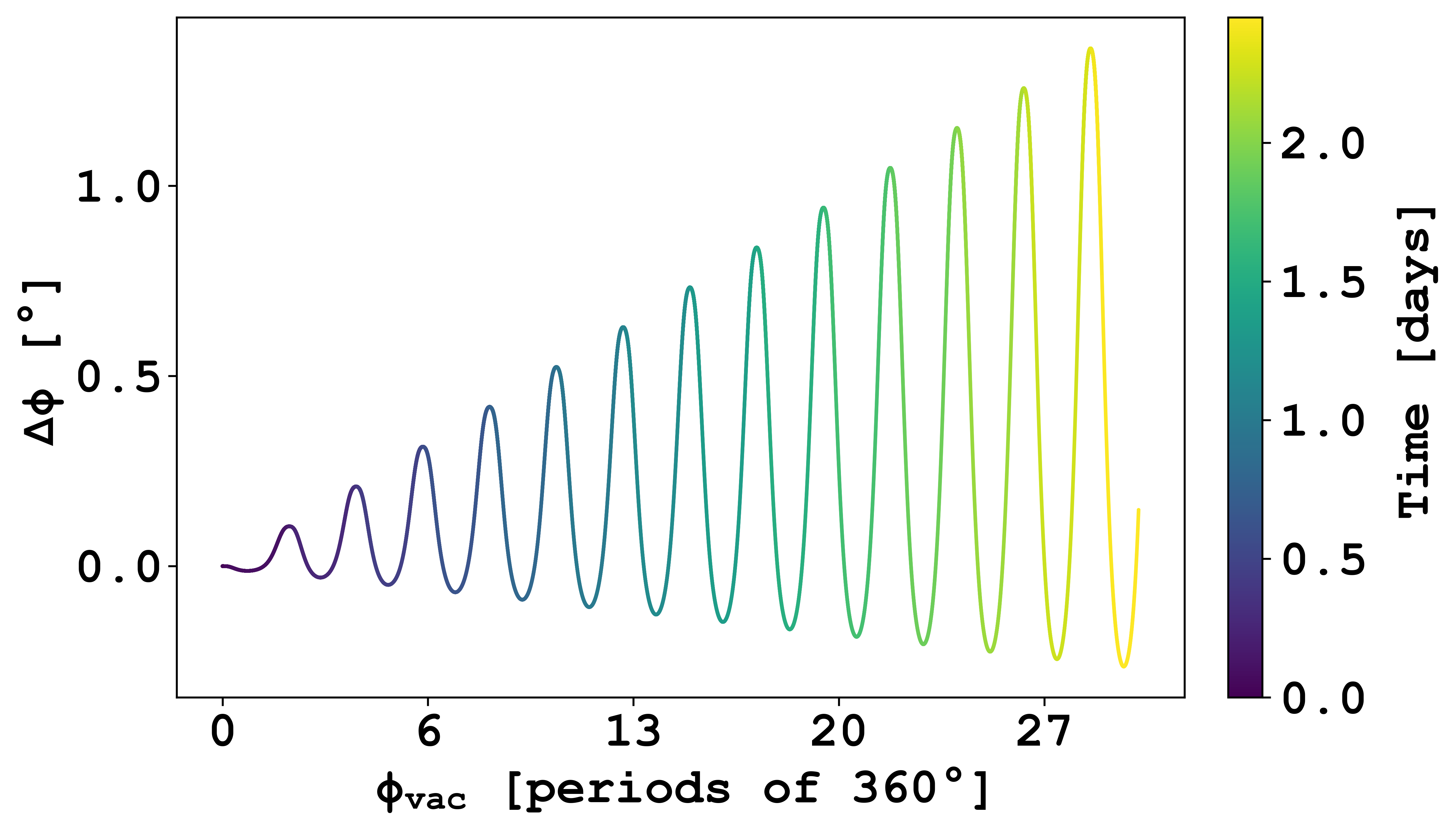}
    \par\smallskip\centering\small Supermassive BH, $a_{\ast}=0.2$ (Kerr \& Compact Cloud).\par
\end{minipage}

\begin{minipage}{0.45\textwidth}
    \centering
    \includegraphics[width=\linewidth]{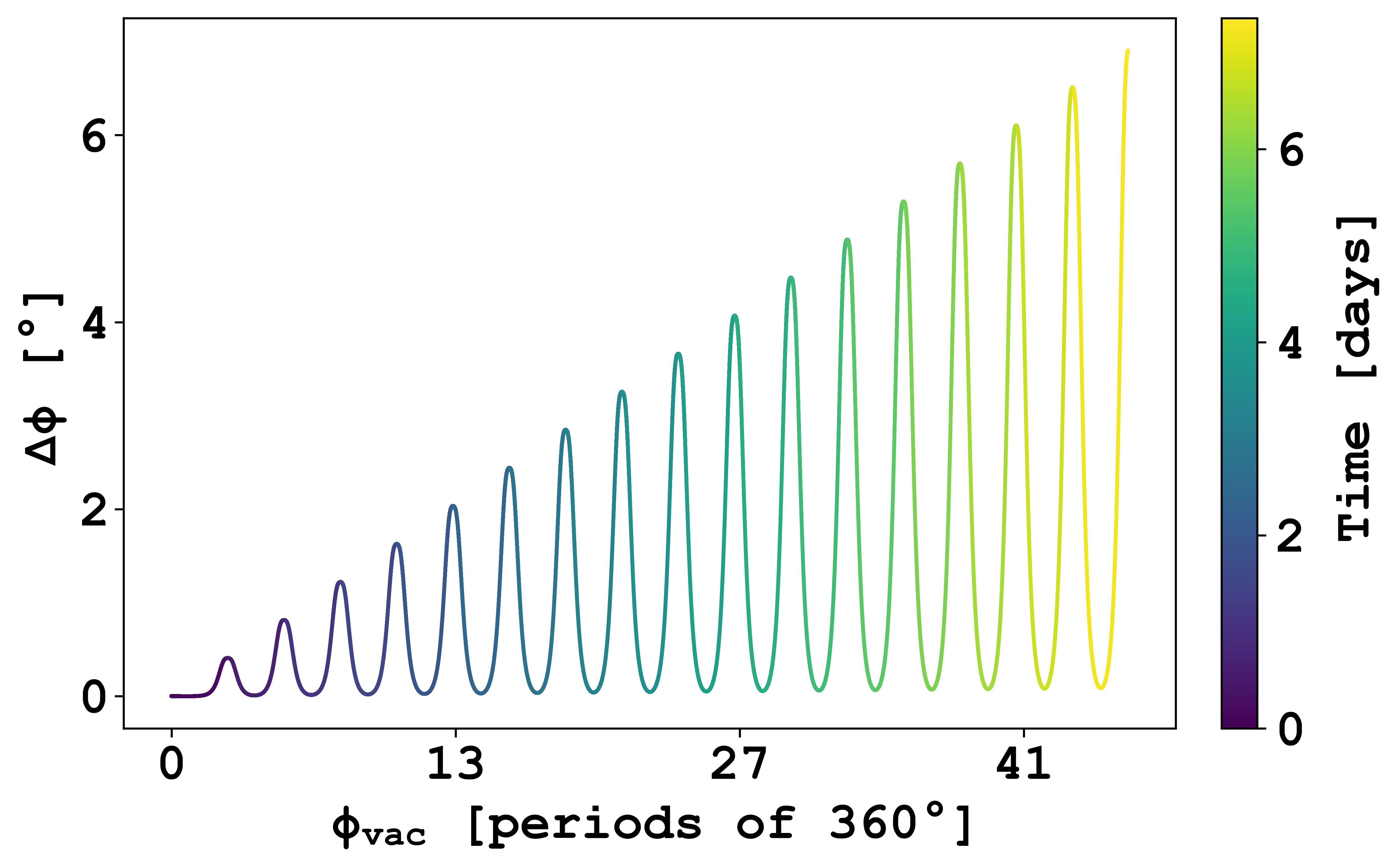}
    \par\smallskip\centering\small Supermassive BH (Schwarzschild \& Galactic Halo).\par
\end{minipage}
\hfill
\begin{minipage}{0.45\textwidth}
    \centering
    \includegraphics[width=\linewidth]{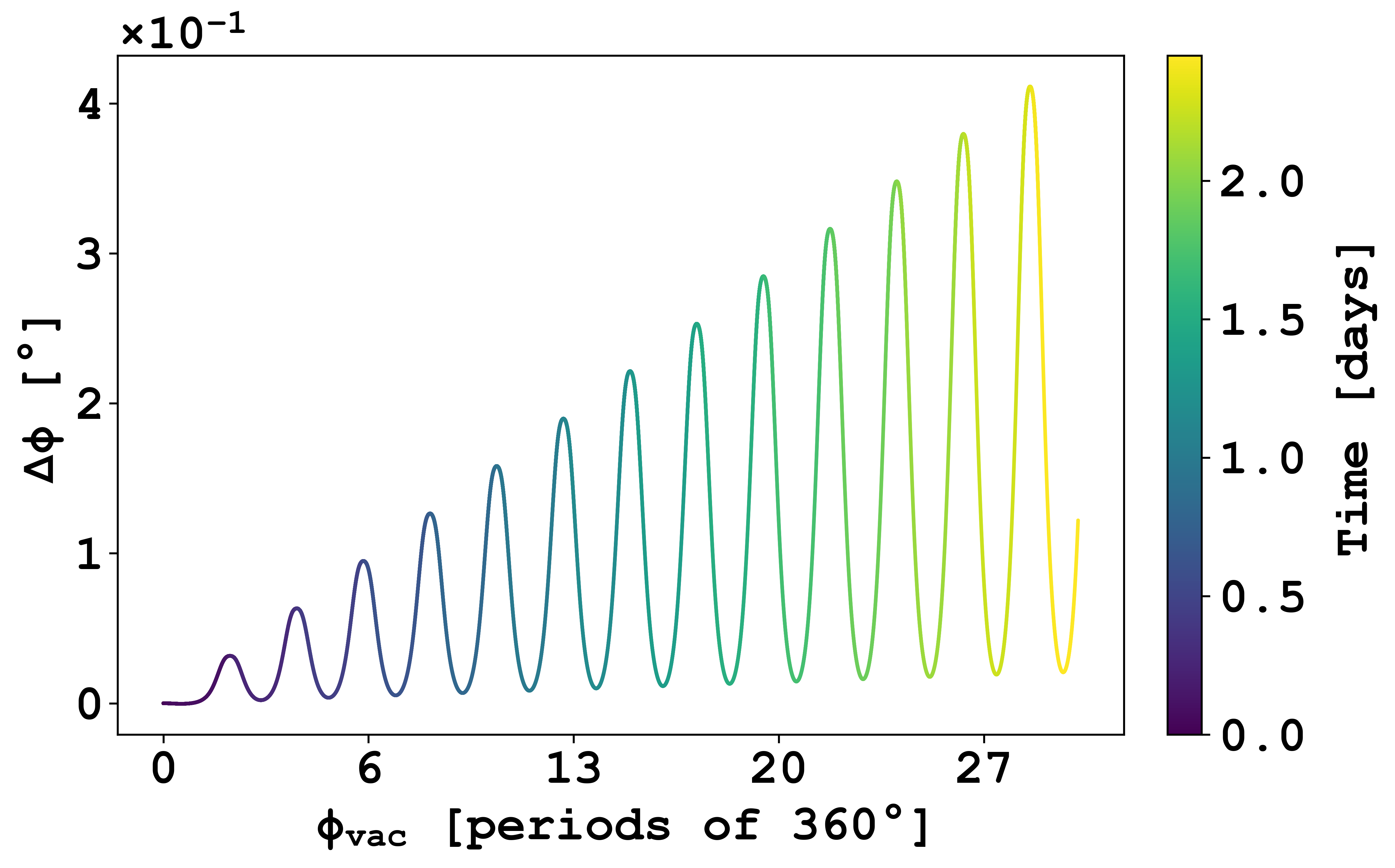}
    \par\smallskip\centering\small Supermassive BH, $a_{\ast}=0.2$ (Kerr \& Galactic Halo).\par
\end{minipage}

\caption{Angular displacement $\Delta_{\phi} = \phi_{\rm SFDM}-\phi_{\rm vac}$ as a function of the corresponding vacuum orbital radius $r_{\rm vac}$ for representative black-hole masses. The left (right) column shows Schwarzschild (Kerr) spacetimes. The first three rows correspond to black holes of $1\,M_{\odot}$, $10\,M_{\odot}$, and a supermassive black hole of $4.3\times10^{6}\,M_{\odot}$, each surrounded by a compact SFDM cloud with total mass $M_{\rm c} = 10^{-5}\,M_{\rm BH}$. The fourth row instead shows the Galactic-center supermassive black hole embedded in a Galactic-scale SFDM halo. The vertical dashed line marks the characteristic radius $r_{\rm char}$, defined from the maximum of $|\Delta r|$ in Fig.~\ref{fig:simulation_block1}. Kerr cases assume a moderate spin $a_{\ast} = 0.2$.}
\label{fig:simulation_block2}
\end{figure*}

For the supermassive black-hole case, we adopt halo parameters representative of the Galactic Center and consider both a Schwarzschild spacetime and a Kerr spacetime with a moderate dimensionless spin parameter, $a_{\ast}=0.2$. The initial conditions adopted in each simulation are summarized in Table~\ref{tab:simulation_parameters} and the results are shown in Fig. \ref{fig:simulation_block1}. For the slowly rotating Kerr black holes considered here ($a_{\ast}\le0.2$), the weak-coupling bound states remain hydrogenic to leading order. Consequently, the characteristic ground-state radius is well approximated by the Schwarzschild expectation value,
$\langle r\rangle_{10}^{(0)}=\frac{3}{2}a_{\rm g}$, with spin-dependent corrections entering only at higher order. Throughout this work, this common leading-order expression is therefore adopted as the analytical reference for both Schwarzschild and Kerr spacetimes. In all cases, the particle is initialized such that its trajectory follows a four-leaf orbit, allowing for a direct comparison between the vacuum and SFDM spacetimes. The characteristic response radius, $r_{\rm char}$, is defined as the location of the maximum radial displacement and is compared throughout this work with the analytical prediction obtained from the hydrogenic expectation value of the bound-state wave function. Complementing the radial response, Fig.~\ref{fig:simulation_block2} shows the accumulated angular displacement.

The correspondence between the perturbed cloud radius and the orbital-response radius has a direct physical origin. The SFDM perturbation mixes the unperturbed hydrogenic states and thereby redistributes the cloud density through $\delta\rho_{\rm cl}(r)$. Because the total cloud mass is conserved, this perturbation primarily changes the radial distribution of the enclosed mass rather than its total value. Consequently, the difference between the perturbed and vacuum gravitational fields is localized over the radial region in which the cloud wavefunction is most strongly redistributed. Test-particle geodesics therefore acquire their largest cumulative radial response while traversing this region.
The characteristic scale of this redistribution is measured by the perturbed expectation value $\langle r\rangle_{n\ell}^{\rm SFDM}$, explaining why the independently determined orbital scale $r_{\rm char}$ tracks the perturbed cloud
radius in the numerical simulations.

As illustrated in Fig.~\ref{fig:radial-phi-sim} of the main text, the correspondence between $r_{\rm char}$ and $\langle r\rangle_{10}^{\rm SFDM}$ persists toward smaller gravitational couplings, where the hydrogenic approximation becomes progressively better controlled.

\newpage

\section{Extending the Perturbative Formalism to Navarro--Frenk--White Halos}
\label{SM:NFW}

Although the main text focuses on scalar-field dark matter halos, the perturbative formalism developed in this work is completely general and can be applied to any weak, static, spherically symmetric dark matter distribution. As an illustration, we derive below the corresponding perturbative corrections for the Navarro--Frenk--White (NFW) halo profile,
\begin{equation}
	\rho_{\mathrm{NFW}}(r)
		=
			\frac{\rho_{\rm s}}{\left(r/r_{\rm s}\right) \left(1+r/r_{\rm s}\right)^{2}}\, ,
	\label{eq:NFW_density}
\end{equation}
where $\rho_{\rm s}$ denotes the characteristic density and $r_{\rm s}$ is the scale
radius of the halo.

The corresponding Newtonian gravitational potential is
\begin{equation}
	\Psi_{\mathrm{NFW}}(r)
		=
			-4\pi\.\.G\.\.\rho_{\rm s}\.\.r_{\rm s}^{2}\,\frac{\ln(1+r/r_{\rm s})}{r/r_{\rm s}}
            \, ,
	\label{eq:NFW_potential}
\end{equation}
where the additive integration constant has been chosen such that $\Psi_{\mathrm{NFW}} \rightarrow 0$ as $r\rightarrow\infty$. Following the same non-relativistic reduction of the Klein--Gordon equation presented previously, we can define the NFW perturbation as
\begin{equation}
	\delta H_{\mathrm{NFW}}(r)
		=
			-4\pi\.\.G\.\.m_{\chi}\.\.\rho_{\rm s}\.\.r_{\rm s}^{2}\;\frac{\ln(1+r/r_{\rm s})}{r/r_{\rm s}}
            \, ,
	\label{eq:dH_NFW}
\end{equation}
where $m_{\chi}$ is the mass of the CDM particle.

Therefore, the first-order energy correction is
\begin{equation}
	\Delta E_{n\ell}
		=
			-4\pi\.\.G\.\.m_{\chi}\.\.\rho_{\rm s}\.\.r_{\rm s}^{2} \int_{0}^{\infty} {\rm d}r\, r^{2} \big\lvert R_{n\ell}(r) \big\rvert^{2}\, \frac{\ln(1+r/r_{\rm s})}{r/r_{\rm s}}\, .
	\label{eq:DE_NFW}
\end{equation}
The new perturbation matrix elements entering the first-order correction to the wavefunction are
\begin{equation}
	\mathcal{V}^\ell_{n'n}
		=
			-4\pi\.\.G\.\.m_{\chi}\.\.\rho_{\rm s}\.\.r_{\rm s}^{2} \int_{0}^{\infty} {\rm d}r\, r^{2} R_{n'\ell}(r)\.\. R_{n\ell}(r)\, \frac{\ln(1+r/r_{\rm s})}{r/r_{\rm s}}\, .
	\label{eq:VNFW}
\end{equation}
Consequently, the correction to the expectation value of the orbital radius is
\begin{equation}
	\Delta\langle r\rangle_{n\ell}
		=
			2\.\.\operatorname{Re} \sum_{n'\.\neq\.n} \frac{ \mathcal{V}^\ell_{n'n}\,\mathcal{R}^\ell_{nn'} }{ E_n^{(0)}-E_{n'}^{(0)} }
            \, ,
	\label{eq:Dr_NFW}
\end{equation}
where
\begin{equation}
	\mathcal{R}^\ell_{nn'}
		=
			\int_{0}^{\infty} {\rm d}r\, r^{3} R_{n\ell}(r)\.\. R_{n'\ell}(r)\, .
\end{equation}
Thus, the NFW halo modifies both the hydrogenic energy spectrum and the characteristic size of the axion cloud through the same perturbative mechanism discussed for the scalar-field dark matter halo. The only difference is the radial dependence of the perturbing potential, which is now determined by the logarithmic NFW profile rather than the oscillatory BEC profile.

\end{document}